\documentclass[twocolumn]{aastex63}

\newcommand{\Msun}{\mbox{M$_{\odot}$}}
\newcommand{\Rsun}{\mbox{R$_{\odot}$}}
\newcommand{\kms} {\mbox{km\,s$^{-1}$}}
\newcommand{\kep}{\textit{Kepler}}
\newcommand{\kic}{KIC\,8736245}

\accepted{2019 September 20}
\submitjournal{The Astronomical Journal}

\shorttitle{Stellar Properties of \kic}
\shortauthors{Fetherolf et al.}

\begin{document}

\title{Stellar Properties of KIC\,8736245: An Eclipsing Binary with a Solar-type Star Leaving the Main Sequence}

\author{Tara Fetherolf}
\altaffiliation{Department of Astronomy, San Diego State University, 5500 Campanile Dr., San Diego, CA 92182, USA}
\affiliation{Department of Physics \& Astronomy, University of California Riverside, 900 University Ave., Riverside, CA 92521, USA}
\author{William F. Welsh}
\affiliation{Department of Astronomy, San Diego State University, 5500 Campanile Dr., San Diego, CA 92182, USA}
\author{Jerome A. Orosz}
\affiliation{Department of Astronomy, San Diego State University, 5500 Campanile Dr., San Diego, CA 92182, USA}
\author{Gur Windmiller}
\affiliation{Department of Astronomy, San Diego State University, 5500 Campanile Dr., San Diego, CA 92182, USA}
\author{Samuel N. Quinn}
\affiliation{Center for Astrophysics ${\rm \mid}$ Harvard {\rm \&} Smithsonian, 60 Garden Street, Cambridge, MA 02138, USA}
\author{Donald R. Short}
\affiliation{Department of Astronomy, San Diego State University, 5500 Campanile Dr., San Diego, CA 92182, USA}
\author{Stephen R. Kane}
\affiliation{Department of Earth and Planetary Sciences, University of California Riverside, 900 University Ave., Riverside, CA 92521, USA}
\author{Richard A. Wade}
\affiliation{Department of Physics \& Astrophysics, The Pennsylvania State University, 525 Davey Lab, University Park, PA 16802, USA}
\correspondingauthor{Tara Fetherolf}
\email{Tara.Fetherolf@gmail.com}

%.............................................................

\begin{abstract} 
There is a well-known stellar parameter discrepancy for late K and M dwarfs, in that the observed radii and temperatures are often respectively larger and cooler than predicted by theory by several percent. In an on-going effort to elucidate this issue, we examine the double-lined \textit{Kepler} eclipsing binary star system KIC\,8736245. We supplement the near-continuous 4-year \textit{Kepler} light curve with ground-based multicolor photometry from Mount Laguna Observatory and spectroscopy from the Hobby-Eberly Telescope. The binary has an edge-on, circular 5.07\,day orbit with stellar masses equal to $0.987\pm0.009$ and $0.782\pm0.009\,\text{M}_\odot$ and radii of $1.311 \pm 0.006$ and $0.804 \pm 0.004\,\text{R}_\odot$, respectively, and an estimated age of 7--9\,Gyr. We find that the stellar radii are consistent with theoretical models within the uncertainties, whereas the temperature of the secondary star is $\sim$6\% cooler than predicted. An important aspect of this work is that the uncertainties derived from a single epoch (individual night of observations) underestimates the overall system parameter uncertainties due to the effect of the \mbox{1--4\%} fluctuations caused by stellar activity. Our error estimates come from the spread in parameters measured at 8 epochs. From the periodicities in the light curve and from the eclipse times, we measure candidate spin periods to be approximately 4.98 and 5.87\,days for the primary and secondary star. Surprisingly, these imply super- and sub-synchronous rotation compared to the orbital period. Thus KIC\,8736245 serves as an interesting case study for the exchange of angular momentum and general stellar astrophysics as stars in binaries evolve off the main sequence. 
\end{abstract}

% This is outdated, but can be used for ArXiv postings
% \keywords{binaries: eclipsing --- stars: activity --- stars: fundamental parameters --- stars: individual (KIC\,8736245) --- stars: rotation --- starspots}
%.............................................................

\section{Introduction}
Accurate measurements of stellar mass, radius, and temperature enable strong tests of stellar evolution theory. Fortunately, eclipsing binary stars provide a direct way to obtain these fundamental stellar parameters, and for the most part the theory and the observations agree. However, several studies have found a discrepancy for main sequence stars with masses $\lesssim$0.8\,\Msun, in that these stars have radii that are $\sim$5--15\% larger and effective temperatures that are $\sim$3--5\% lower than predicted by theoretical models \citep[e.g.,][]{Torres02,Ribas06,Torres06,Lopez-Morales07,Ribas08,Torres10, Morales10,Feiden12,Spada13,Torres13}. The disagreement in radius and temperature for these low-mass stars have typically been attributed to enhanced magnetic fields that block convection, producing star spots on the surface that lowers the surface temperature and ``bloats'' the radius of the star \citep[e.g.,][]{Ribas06,Torres06,Chabrier07,Lopez-Morales07,Kraus11}. Enhanced star spot activity is especially expected in short-period binaries ($P<10$\,days), since magnetic activity is enhanced by the faster spinning stars that are tidally locked to their orbital periods \citep{Ribas06, Torres13, Lurie17}. Most well-studied low-mass eclipsing binaries have orbital periods shorter than 3\,days \citep{Devor08-1,Torres13}, but unlocking the nature of the discrepancies in the radius and temperature of low mass stars requires observations of low-mass stars in systems with longer orbital periods that are not necessarily tidally synchronized.

Accurate masses are best measured from double-lined spectroscopic binaries, while measuring accurate radii generally requires that the system is eclipsing \citep[see][]{Torres10}. However, double-lined eclipsing binaries with one or more low-mass stars are rare since the secondary star is usually too faint to spectroscopically observe from the ground. Furthermore, most ground-based surveys cannot continuously observe for more than a few hours at a time and consequently miss eclipses of long-period systems. The \kep\ space telescope \citep{Borucki10, Koch10} provided near continuous observations of $\sim$170,000 stars with the goal of discovering Earth-size planets around Sun-like stars, but also found other exciting astrophysical objects---including identifying $\sim$3000 eclipsing binary systems\footnote{\url{http://keplerebs.villanova.edu/}} \citep{Prsa11, Slawson11, Kirk16}. The continuous, high-precision photometry also made possible the careful study of periodic signatures in the light curves, such as pulsations, phase variations, and stellar activity---especially for systems with long orbital periods \citep[e.g.,][]{Lurie17}. 

In this paper we measure the orbital parameters of \kic\ \citep[$\alpha = 18^h53^m44.179^s$, $\delta = +44^{\circ}59'23.03''$, J2000, $r=13.8$;][]{Brown11}, a double-lined eclipsing binary with a $\sim$5\,day circular orbital period consisting of a Solar mass primary and $\sim$0.8\,\Msun\ secondary star \citep{Devor08}. This system provides a good test of stellar theory given its intermediate orbital period, secondary star being near the upper limit for low-mass stars with observed inflated radii, and evidence for magnetic activity in its light curve. 

% UPDATE if organization changes.
In \autoref{sec:data} we present our photometric and spectroscopic data. In \autoref{sec:elc} we solve for the orbital parameters of \kic\ using the Eclipsing Light Curve package \citep[ELC;][]{Orosz00}. We examine and discuss the periodic signatures identified in the out-of-eclipse portions of the \kep\ light curve and the eclipse timing variations in \autoref{sec:periodicity}. Finally, we summarize our results in \autoref{sec:summary}.
%
%
%

% ................................
\section{Observations and Data Processing}\label{sec:data}
% NO CONTENT HERE

\subsection{Space-based Photometry}\label{sec:kepler}
\begin{figure*}
\epsscale{1.15}
\plottwo{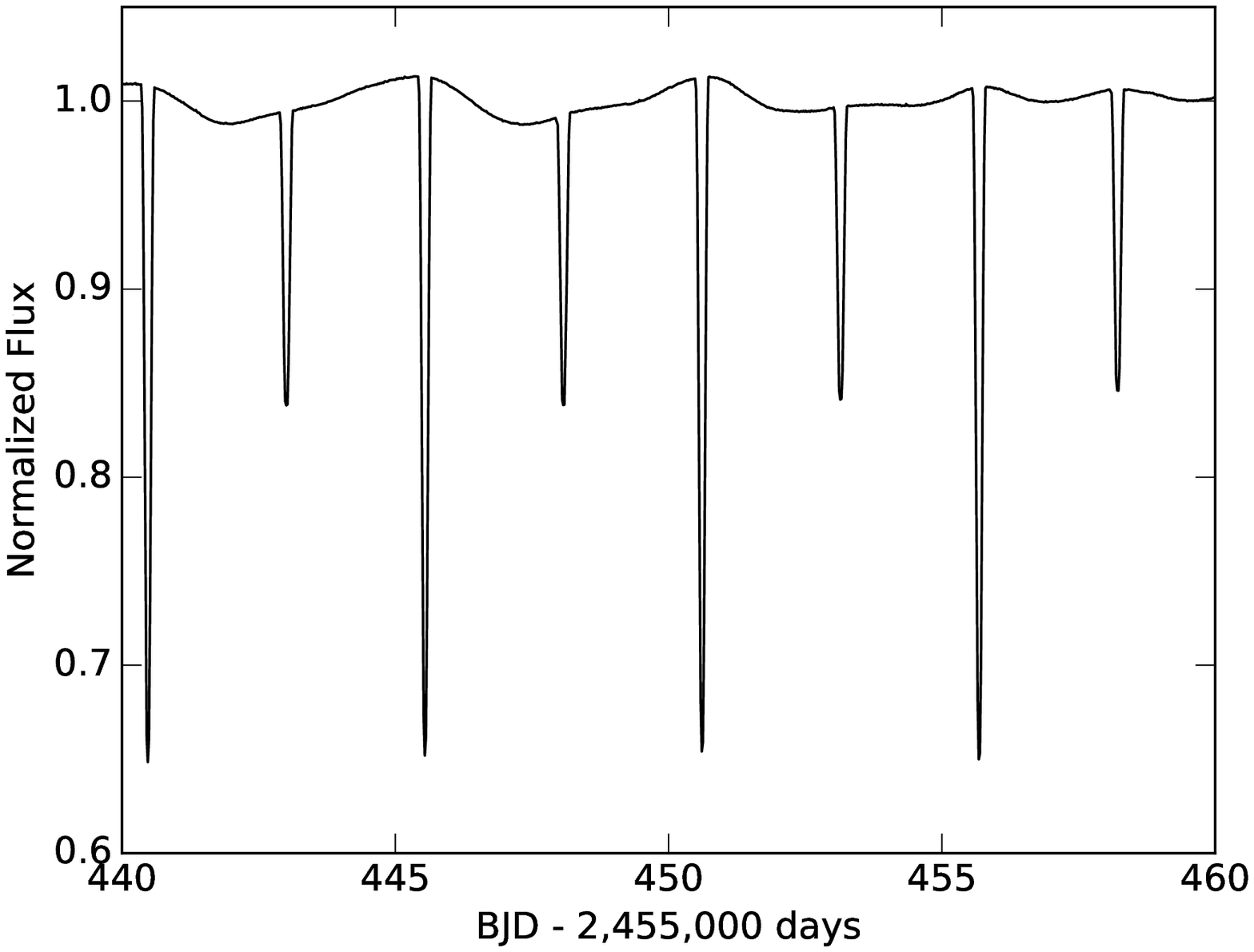}{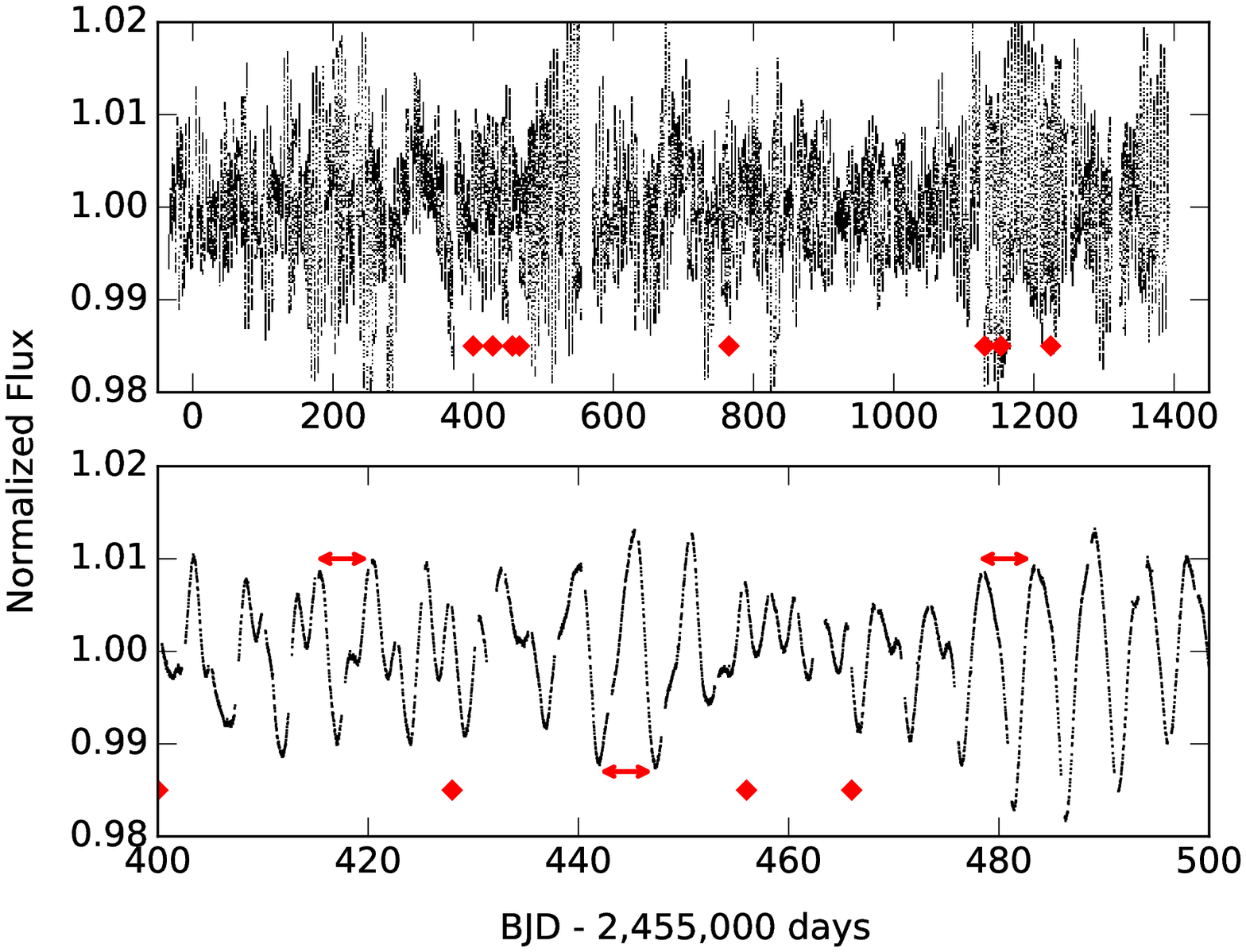}
\caption{\textit{Left:} Normalized \kep\ light curve of \kic, showing a 20\,day segment from Q6. \kic\ exhibits deep 35\% primary eclipses and 16\% secondary eclipses at a 5.07\,day orbital period.
\textit{Top right:} Out-of-eclipse normalized \kep\ light curve for Quarters 1--16. \kic\ experiences 1--4\% peak-to-peak quasi-periodic fluctuations in stellar brightness, attributed to stellar activity. \textit{Bottom right:} Zoomed-in region of the top panel, showing a 100\,day time segment from Q6 and Q7. The red diamonds mark the dates when ground-based photometry was obtained from Mount Laguna Observatory. The arrowed lines denote the duration of one full orbital period ($P_{\text{orb}}=5.07$\,days).}
\label{fig:lcspots}
\end{figure*}
Nearly continuous, high-precision long-cadence (29.4\,minutes) photometric observations were taken of \kic\ by the \kep\ space telescope \citep{Borucki10, Koch10} from March 2009 through May 2013. We obtain simple aperture photometry \citep[SAP;][]{Jenkins10-1, Jenkins10} light curves from the Mikulski Archive for Space Telescopes\footnote{\url{http://archive.stsci.edu/kepler/}} (MAST) \kep\ Data Release 24 for Quarters (Q) 1--16, which covers 1426\,days of near continuous observations\footnote{We did not include Q17. It contains five primary and secondary eclipses, compared to the 229 primary and 230 secondary eclipses observed throughout Q1--16.}. Only observations with Data Quality flag less than 16 were used, and in total data gaps were present only 9\% of this time, occurring during monthly spacecraft data downloads, safe modes, and any other spacecraft or photometer anomaly. 

To preserve the out-of-eclipse fluctuations attributed to star spot activity, we apply our own detrending technique as described by \citet{Bass12}. Briefly, the SAP data is detrended in small sections, separated by any observation breaks that were $>$1\,day in duration. Then an $n$-piece cubic spline interpolating function (where $n$ is typically 10--30) is fit to the out-of-eclipse portions of the light curve and is used to normalize each subsection. 

A 20\,day segment of the \kep\ light curve is shown in the left panel of \autoref{fig:lcspots}. The photometry reveals \kic\ to exhibit deep 35\% primary and 16\% secondary eclipses at a 5.07\,day orbital period. The right panel of \autoref{fig:lcspots} shows the out-of-eclipse light curve from Q1--16 in the top panel and a 100\,day segment from Q6--7 in the bottom panel. The 1--4\% peak-to-peak quasi-periodic fluctuations in stellar brightness that are approximately equal to the orbital period are attributed to spots on the stars rotating into and out of the line-of-sight. 

\begin{figure*}
\epsscale{1.15}
\plottwo{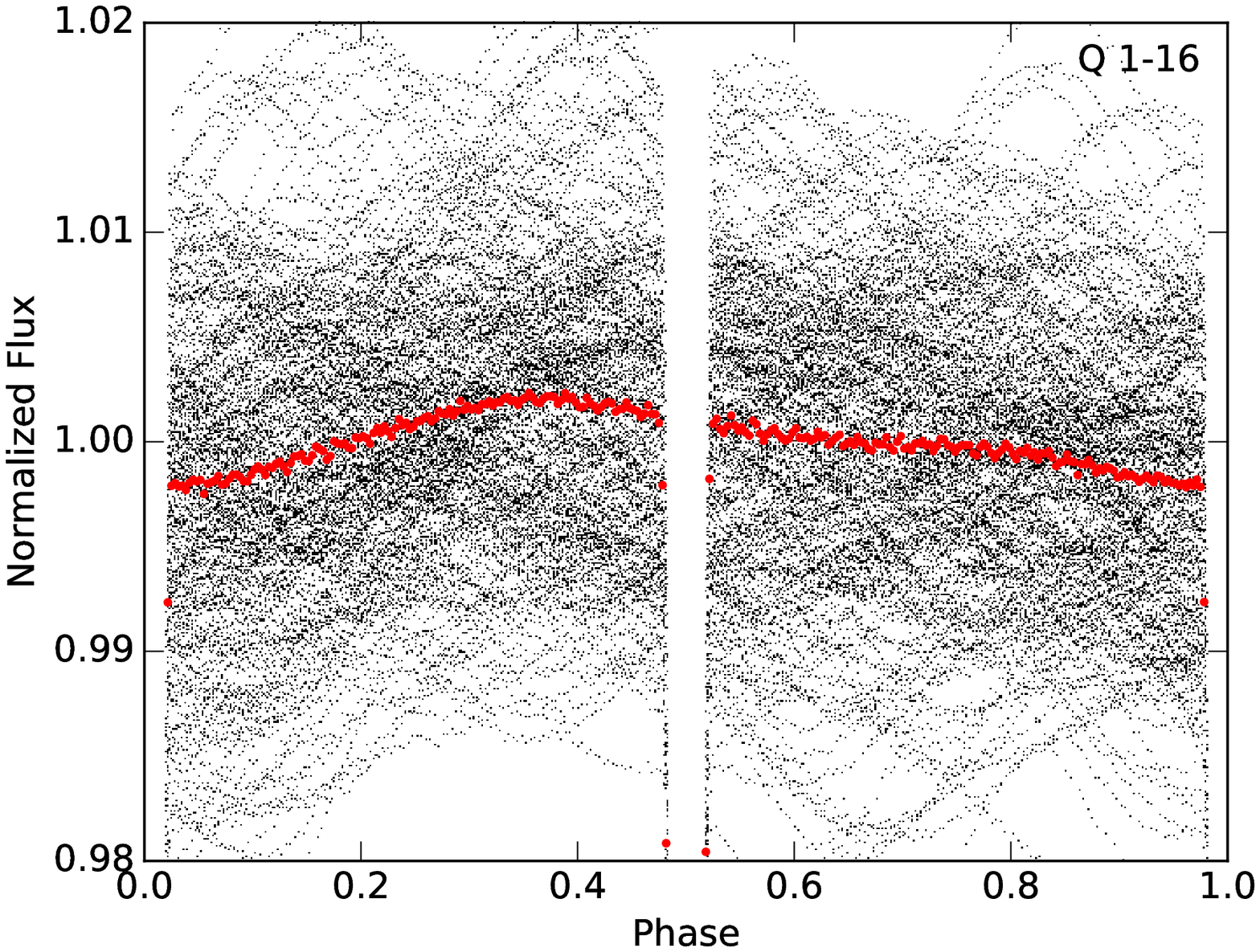}{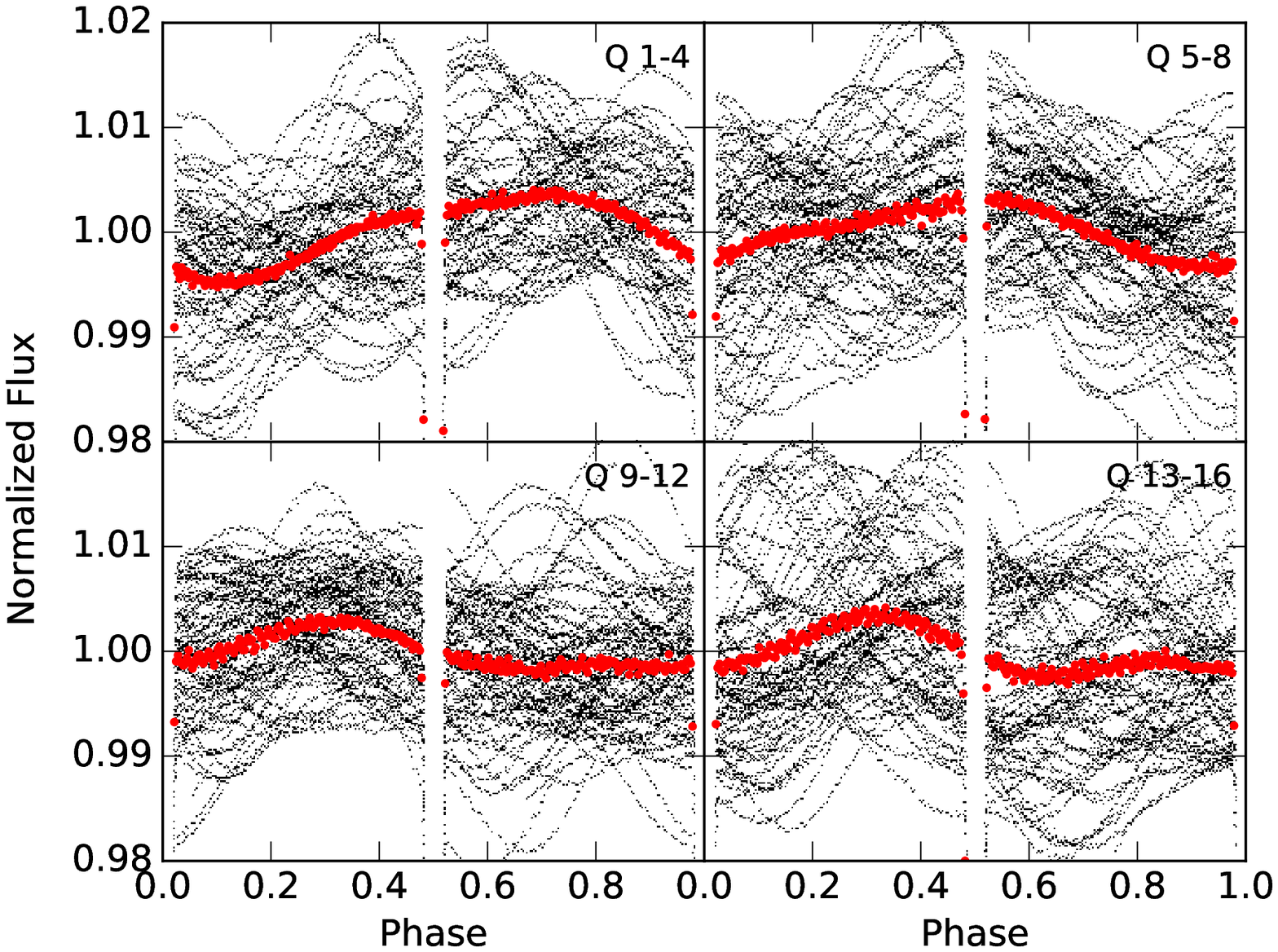}
\caption{The \kep\ light curve phase folded and averaged into 300\,bins for Quarters Q1--16 (\textit{left}) and in 1\,year segments (\textit{right}). The black points are the individual \kep\ data points and the red circles are the averages per bin. The non-repeatability shown in the yearly figures indicates that the modulations due to stellar activity do not average out over this timescale.}
\label{fig:lcphase}
\end{figure*}

Due to stellar activity, fluctuations on a timescale of days are present in the light curve. We investigate whether these variations average out on long timescales by folding and averaging the \kep\ photometry into 300\,bins. The bins are overlain on the folded light curve in the left panel of \autoref{fig:lcphase}, and show that the out-of-eclipse regions are not flat even after averaging together 4\,years of observations. The residual signal could be caused by periodic orbital modulations induced by gravitational and geometric effects, such as Doppler beaming, ellipsoidal variations, and reflection effects \citep[e.g.,][]{Faigler11}. While the $\sim$0.002\% residual modulations in the left panel of \autoref{fig:lcphase} appear to be consistent with ellipsoidal variations, the right panels of \autoref{fig:lcphase} show that the shape of the residuals is not consistent between smaller, 1-year segments of time. Therefore, we conclude that the out-of-eclipse signal in the binned data is a result of star spot variations that do not average out. 

\subsection{Ground-based Photometry}\label{sec:mlo}
Multi-waveband ground-based photometry was obtained on 8 separate nights between 2010\,July\,22 and 2012\,October\,23 (UT) using the 0.6\,m and 1.0\,m telescopes at Mount Laguna Observatory (MLO; located near San Diego, California). Observations on the 0.6\,m telescope were taken with a SBIG\,STL-1001E CCD camera in Johnson/Bessel $V$, $R$, and $I$ filters and observations on the 1\,m telescope were taken with a $2048\times2048$\,pixel 2005 Lorel CCD camera in Kron $B$ and $V$ filters. Exposures were 1--2 minutes long depending on the seeing, and typically the center and either the ingress or egress were obtained due to the eclipse duration of 5.475\,hours compared to the short summer nights when the \kep\ field was visible. Each night we obtained twilight sky flats, and took dark current images at the same CCD operating temperature and exposure length as the science images. Standard calibrations are applied using the Astronomical Imaging Processing software (AIP4Win version 2.4.1) created by Richard\,Berry and James\,Burnell (published by Willmann-Bell\,Inc.). Together, the auto-calibration and multi-image photometry tools follow standard treatment of darks and flats to perform differential magnitude photometry. 

As is typical of ground-based relative photometry, the MLO light curves often include systematic tilts. These artifacts of imperfect calibration require careful removal such that the intrinsic tilt caused by the stellar activity remains. To retain the intrinsic tilt caused by star spot modulations, we match each MLO light curve with its simultaneously observed \kep\ light curve. The tilt is removed by fitting a cubic polynomial over the out-of-eclipse portions of each respective \kep\ light curve, then subtracting the polynomial from the MLO photometry. The MLO light curve is then normalized to be consistent with the \kep\ photometry by offsetting the MLO data by the difference between the cubic polynomial that was fit to the \kep\ photometry and the mean out-of-eclipse flux of the MLO light curve. 

During three of the ground-based observations (2010\,July\,22, 2011\,July\,22, and 2012\,July\,21), the equatorial mount on the MLO 0.6\,m telescope required a pier flip as the observed target crossed the meridian. After a pier flip, the image focal plane rotated $180^\circ$ and the stars fell on a different set of CCD pixels. Ideally the flat fielding calibration would account for the different pixel sensitivities so that the pre- and post-pier flip light curves connect together smoothly, but this is not always the case (perhaps due to imperfect flat fielding caused by scattered light). On the nights of 2010\,July\,22, 2011\,July\,22, and 2012\,July\,21 an addition normalization (magnitude offset) step was required. The standard calibration procedure was applied separately to observations on each side of the pier flip, and systematic tilts were removed from the individual light curve segments as described above. The resulting two separate light curves were each separately fit with the ELC code (see \autoref{sec:elc}) along with the corresponding \kep\ data and radial velocities, simply to determine their relative magnitude offset. This offset was then applied to the post-pier flip light curve. The process was repeated a second time to refine the offset, and then the data segments were combined into one light curve for the night.

\subsection{Spectroscopy}\label{sec:spectra}
\begin{deluxetable}{lrr}
\tablecaption{\kic\ Radial Velocity Measurements \label{tab:spec_data}}
\tablehead{\colhead{Time} & \colhead{RV$_1$} & \colhead{RV$_2$} \\ [-5pt] \colhead{(BJD-2455000)} & \colhead{(\kms)} & \colhead{(\kms)}}
\startdata
341.77552 & $24.49\pm0.51$ & $-1.54\pm1.13$ \\
342.79340 & $77.89\pm0.68$ & $-70.60\pm1.28$ \\
347.77079 & $76.50\pm0.28$ & $-69.16\pm0.55$ \\
349.76928 & $-29.23\pm0.17$ & $65.00\pm0.35$ \\
372.70794 & $61.61\pm0.20$ & $-49.49\pm0.25$ \\
426.77896 & $-50.68\pm0.20$ & $91.78\pm0.37$ \\
458.69597 & $49.32\pm0.37$ & $-34.26\pm0.80$ \\
472.65980 & $-41.45\pm0.11$ & $80.25\pm0.23$ 
\enddata
\end{deluxetable}
Echelle spectroscopy was obtained on 8 nights between 2010\,May\,25 and 2010\,October\,3 (UT) using the High Resolution Spectrograph \citep[HRS;][]{Tull98} on the 10\,m Hobby-Eberly Telescope \citep[HET;][]{Ramsey98}. The HRS instrument was configured to a 30,000 resolving power, the central echelle rotation angle, the 2\arcsec \ science fiber, two sky fibers, and the 316\,groove\,mm$^{-1}$ cross disperser was set to give a central wavelength of 6948\,\AA. To aid the removal of cosmic rays, the 600\,s exposure times were split into two parts of 300\,s each. We subtract the electronic bias from each image and remove cosmic rays using \texttt{crreject} in IRAF. Then the ``blue" echelle spectra (wavelength coverage 5100--6900\,\AA) are extracted and wavelength calibrated using the IRAF \texttt{echelle} package.

Radial velocities are measured using the ``broadening function'' technique \citep{Rucinski92, Rucinski02}, which is well-suited for double-lined spectroscopic binaries with component velocity separations that are on the order of the spectral resolution. Refer to \citet{Bayless06} for a detailed discussion of applying the broadening function technique to HET spectra. The broadening functions exhibit two significant peaks, indicating that the system is a double-lined spectroscopic binary. A third peak is also observed, but its strength is dependent on the phase of the Moon and it remained consistent with the systemic velocity over 4\,months of observations. Therefore, the third peak is deemed unrelated to the system. The measured radial velocities measurements are listed in \autoref{tab:spec_data} and shown in the left panel of \autoref{fig:kepecl}. The radial velocities indicate that the system has a nearly circular orbit, which is consistent with the secondary eclipse being precisely at 0.5\,phase. 

We measure the effective temperatures ($T_\mathrm{eff}$), surface gravities ($\log{g}$), projected rotational velocities ($v\sin{i}$), and metallicity ([m/H]) of the stars from the spectra using a method similar to the one described in \citet{Kostov16}. We use the two-dimensional cross-correlation technique TODCOR \citep{Zucker94} with template spectra from the CfA library of synthetic templates \citep[see, e.g.,][]{Nordstroem94, Latham02} to assess the best-fit stellar parameters for both stars simultaneously. We carry out the analysis using 210\,\AA\ of the spectrum between 5150--5360\,\AA, which includes the gravity sensitive Mg I b triplet near 5190\,\AA\ and overlaps with the wavelength range of the CfA templates. For every pair of templates in the range $4500\,{\rm K} < T_\mathrm{eff} < 6500\,{\rm K}$, $3.5 < \log{g} < 5.0$, and $-1.5 < {\rm [m/H]} < +0.5$, we calculate the mean cross-correlation peak height across all eight spectra. We then interpolate to the peak of the surface defined by these cross-correlation coefficients and adopt the stellar parameters corresponding to the location of the peak. The spectroscopic parameters suffer from strong degeneracies such that relatively large changes in one parameter can be compensated by changes in the others in order to maintain a good fit to the observed spectrum. To partially overcome this degeneracy, we interpolate to the surface gravities determined from the light curve and radial velocity analyses, and find $T_\mathrm{eff,1} = 5810 \pm 100$\,K, $T_\mathrm{eff,2} = 5030 \pm 125$\,K, and ${\rm [m/H]} = -0.31 \pm 0.15$. However, these internal uncertainties do not account for remaining correlated errors in the primary and secondary $T_\mathrm{eff}$\ and [m/H], which we measure to be 70\,K, 28\,K, and 0.08\,dex, respectively. The errors reported in \autoref{tab:spec_results} represent the combined internal and correlated errors. Finally, we note that we fit $v\sin{i}$ iteratively for computational considerations, and because it does not suffer from the same degeneracies as the other three parameters. All measured spectroscopic properties are listed in \autoref{tab:spec_results}.
\begin{deluxetable}{lc}
\tablecaption{\kic\ Spectroscopic Properties \label{tab:spec_results}}
\tablehead{\colhead{Parameter} & \colhead{Value}}
\startdata
$K_1$ (\kms) & $66.17\pm0.2$ \\
$K_2$ (\kms) & $83.48\pm0.2$ \\
$e$ & 0 \\  % ??
$\gamma$ (\kms) & $12.45\pm0.1$ \\
$T_{\text{eff,1}}$ (K) & $5810\pm120$ \\
$T_{\text{eff,2}}$ (K) & $5030\pm130$ \\
$\log{g}_1$ (dex) & 4.189 (fixed) \\
$\log{g}_2$ (dex) & 4.530 (fixed) \\
$v\sin{i}_1$ (\kms) & $14.7\pm2.0$ \\
$v\sin{i}_2$ (\kms) & $8.5\pm1.5$ \\
$\mathrm{[m/H]}$ (dex) & $-0.31\pm0.17$
\enddata
\end{deluxetable}

\section{Light Curve Modeling}\label{sec:elc}
We use the Eclipsing Light Curve code \citep[ELC;][]{Orosz00, Wittenmyer05} to model the light curves and radial velocities. For this binary, we use ELC in its ``numerical'' mode, which covers the stellar surfaces with specific intensity tiles, then sums them. While much slower than the ``analytic'' mode, this mode allow us to include simple star spots. A Markov chain Monte Carlo algorithm is used to determine the model parameter values and uncertainties. In total, we fit for 26 free parameters, although not all simultaneously as we explain below. The orbital period $P$ ($= 5.069482\pm0.000001$\,days) is well constrained by the $\sim$230\,primary and secondary eclipses in the \kep\ light curve, as is the time of conjunction $T_{\text{conj}}$. We fit for the primary mass $M_1$, the mass ratio $q=M_2/M_1$, the primary radius $R_1$, and the radius ratio $R_1/R_2$. The temperature of the primary star is fixed at the spectroscopically determined value of 5810\,K, but the temperature of the secondary star is allowed to vary within the spectroscopic uncertainties. This allows the multi-color ground-based photometry to help constrain the temperature ratio. We also fit for the inclination $i$, eccentricity $e$, argument of periastron $\omega$, and the \kep\ background light contamination, which is estimated to be 2.2\%  according to the MAST archive. Finally, the modulations in the light curve are modeled by two star spots on each star. Each spot is specified by four parameters: angular size, temperature ratio relative to the star's surface temperature, and its location in latitude and longitude. These star spots do not evolve with time and so only short segments of the light curve can be used.

% UPDATE if subsections are adjusted.
We discuss our initial fitting procedure using the \kep\ eclipses and radial veloctities in \autoref{sec:initmod}, then we incorporate our ground-based photometry into the ELC fitting in \autoref{sec:nightmod}. Finally, we estimate and present the best-fit orbital parameters in \autoref{sec:finparams}.

\subsection{Initial Orbital Parameter Estimates}\label{sec:initmod}
\begin{figure*}
\epsscale{1.15}
\plottwo{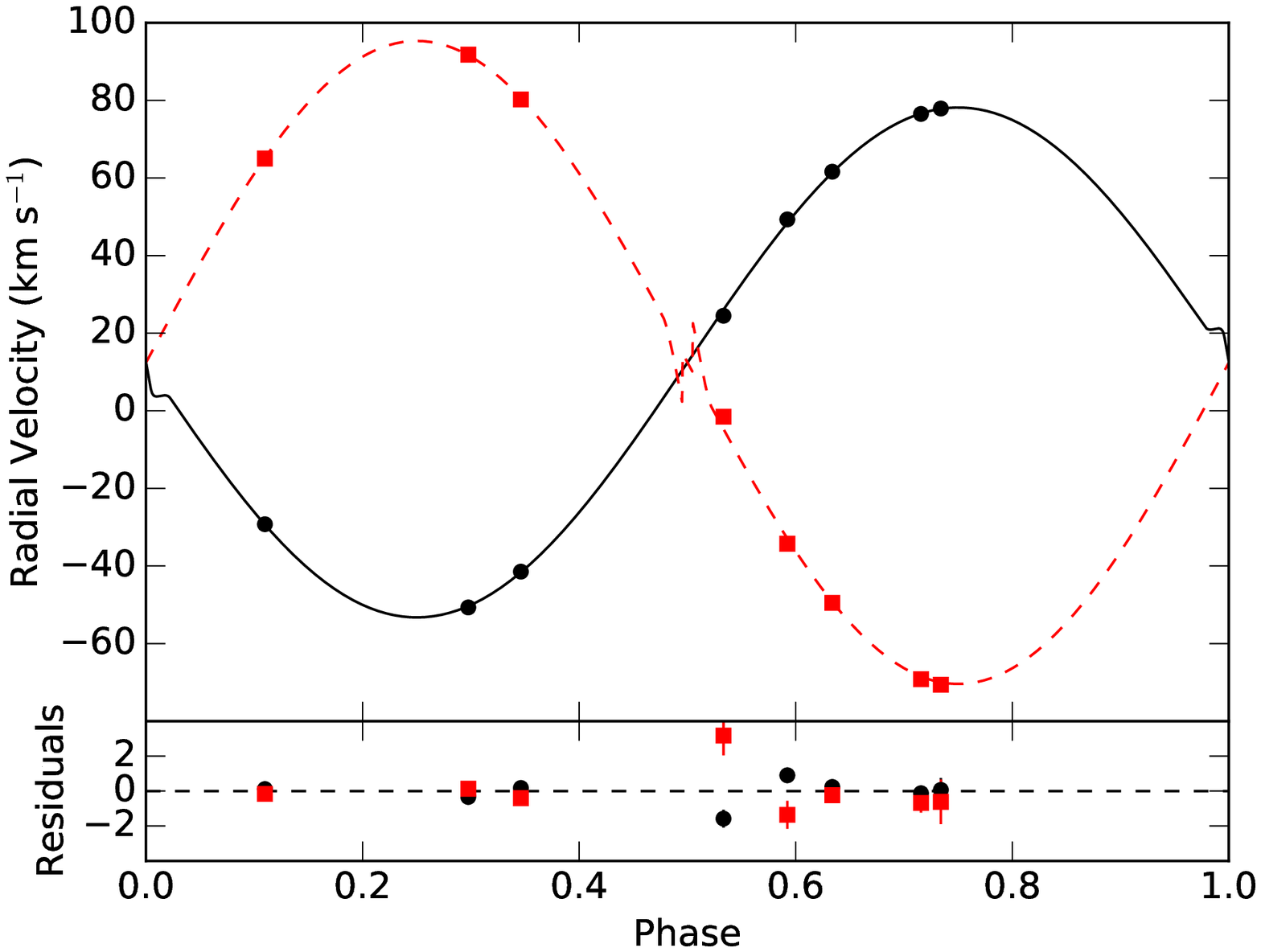}{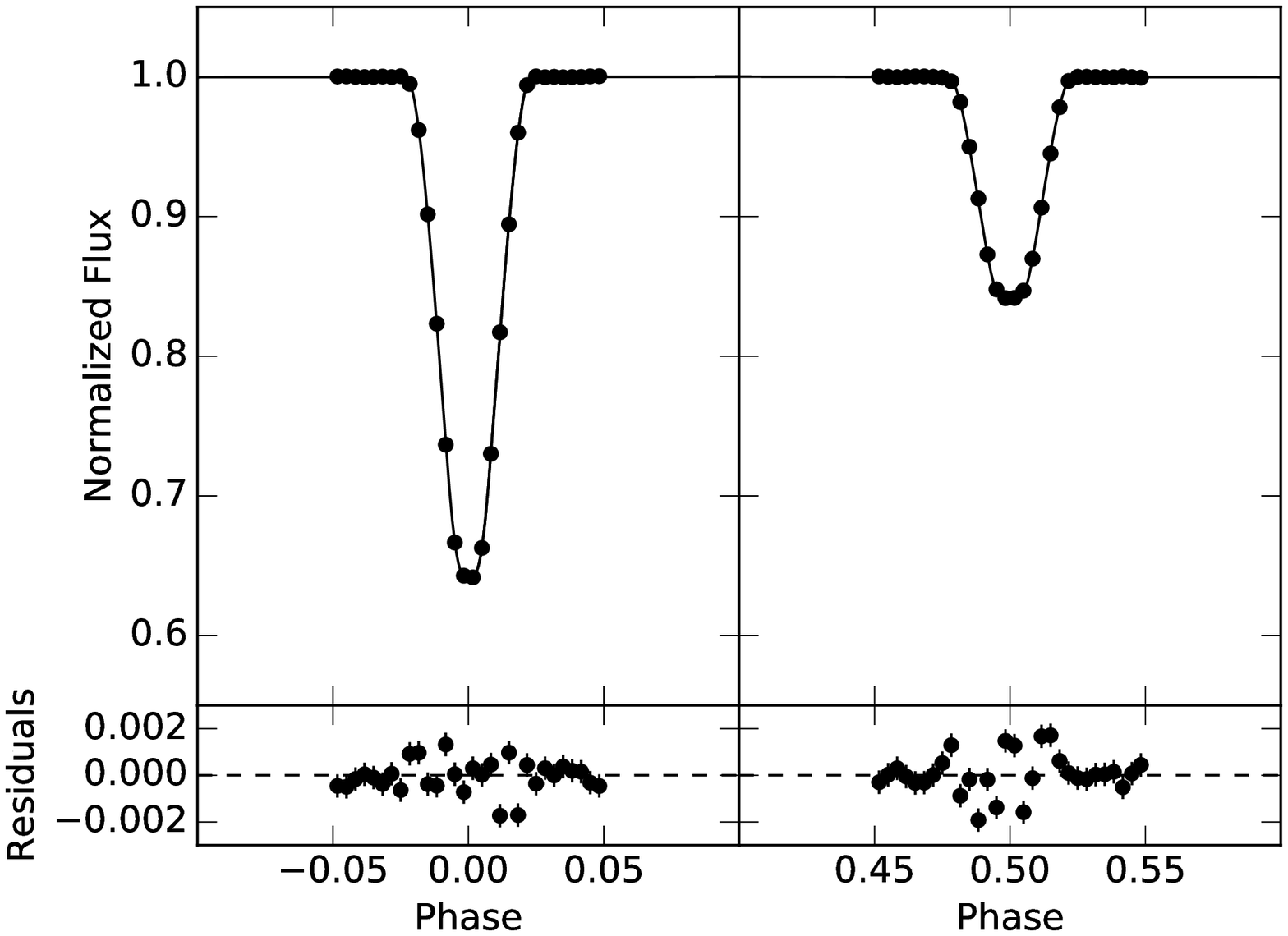}
\caption{\textit{Right:} Radial velocities and the best-fit ELC model, with residuals shown in the lower panel. The black and red colors represent radial velocities of the primary and secondary stars, respectively. \textit{Left:} The orbit phase-folded and binned \kep\ light curve, after removing any out-of-eclipse tilts. The solid curve is the best-fit ELC model. The errors have been set to $\pm$0.0005\,magnitudes so that each bin is evenly weighted during the modeling.}
\label{fig:kepecl}
\end{figure*}
It is clear from \autoref{fig:lcspots} that star spots are important in this binary and can bias the system parameters if ignored, but including star spots in the modeling presents challenges. Modeling the star spots is both computationally expensive, and, since our model does not allow for star spot  evolution, only short segments of the data (approximately the length of the stellar rotation period) can be fit. To overcome these obstacles, we split the modeling process into two parts: solving for spot-independent parameters, then spot-dependent parameters. The first part uses all of the \kep\ data, the second part uses individual epochs and the ground-based photometry. Both use the radial velocity measurements.

\begin{figure}
\epsscale{1.15}
\plotone{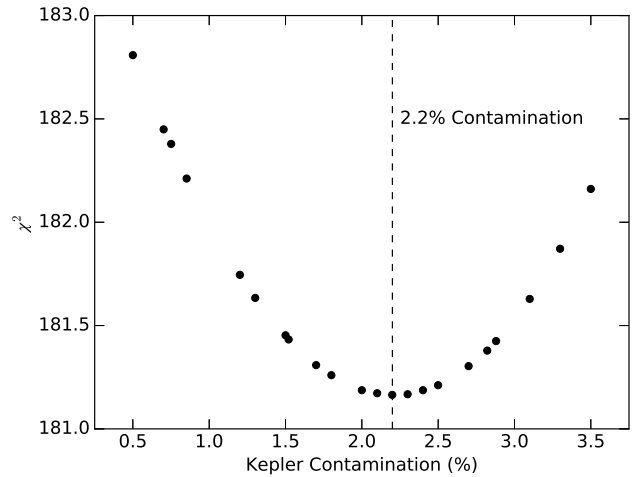}
\caption{Chi-squared versus the preset contamination level, where all other parameters were set free in the ELC modeling of the phase folded and binned \kep\ light curve. The lowest chi-squared is consistent with 2.2\% contamination, which is in agreement with the contamination level stated on the MAST archive.}
\label{fig:cscont}
\end{figure}

For the first step, we phase fold and bin the \kep\ light curve, and discard any data more than $\pm$0.05\,phase away from the eclipses. We remove any tilt by dividing by a linear function fit through the out-of-eclipse portions of the light curve. We also set the uncertainties to 0.0005 ($\sim$3 times their mean value), so that each bin receives equal weighting. The folded, binned, and trimmed light curve is shown in the right panels of \autoref{fig:kepecl}. This simplified light curve allows us to use the fast analytic mode in ELC in order to estimate the system parameters. We solve for the masses, radii, temperatures, eccentricity, argument of periastron, and \kep\ contamination. The latter three will then be held fixed in the individual epoch fitting. We fix the \kep\ contamination to several constant values between 0.5--3.5\% while allowing all other parameters to be free. In \autoref{fig:cscont} we plot the $\chi^2$ of the model versus the contamination, clearly showing a minimum at 2.2\%. This is in excellent agreement with the estimated contamination level stated on the MAST archive, so we fix the contamination to be 2.2\% for all subsequent models. We find the eccentricity to be $<$10$^{-6}$, therefore it is fixed to zero (and the argument of periastron becomes irrelevant). The best-fit model is shown in \autoref{fig:kepecl}. The extra noise in the residuals near the eclipse centers is likely due to star spot crossing events that do not cancel out in the averaged \kep\ light curve. However, we reiterate that this initial fit is simply used to quickly find appropriate starting parameters for the much slower individual epoch fitting on the simultaneously observed \kep\ and MLO photometry (see \autoref{sec:nightmod}).

\subsection{Orbital Parameter Measurements by Epoch}\label{sec:nightmod}
\begin{figure*}
\plotone{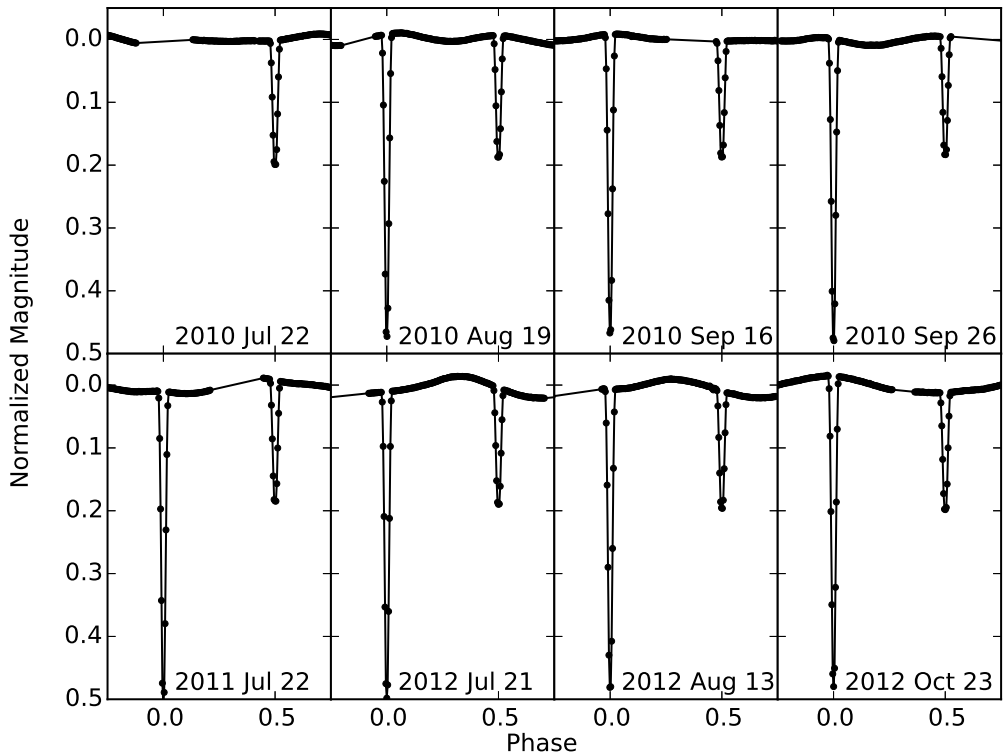}
\caption{The \kep\ photometry associated with each ground-based MLO observation. Notice the modulations in the out-of-eclipse sections of the light curve, due to star spots. The black curves shows the best-fit ELC model to the light curves. Note that there is no primary eclipse data on 2010\,Jul\,22 due to a gap in the \kep\ observations.}
\label{fig:nightkep}
\end{figure*}
\begin{figure*}
\plotone{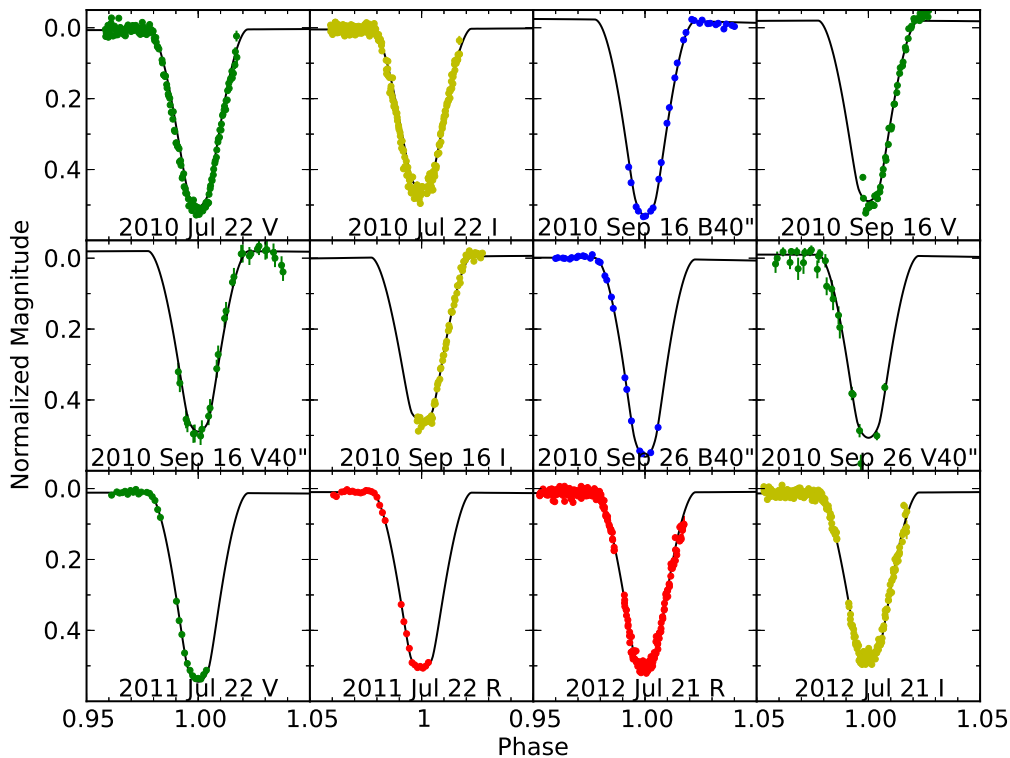}
\caption{Primary eclipse light curves obtained from MLO observations in $B$ (blue points), $V$ (green points), $R$ (red points), and $I$ (yellow points) bands. The black curves show the ELC model that best-fits the photometry. Most observations were obtained using the 0.6\,m telescope, unless otherwise noted following the filter label (i.e. 40''). Note that 2010\,Sep\,16 has two $V$-band observations due to simultaneous observations taken on the 0.6\,m and 1.0\,m MLO telescopes. The out-of-eclipse light has been matched to the \kep\ light curve and thus has not been renormalized to an instrumental magnitude of zero. }
\label{fig:lcprim}
\end{figure*} 
\begin{figure*}
\plotone{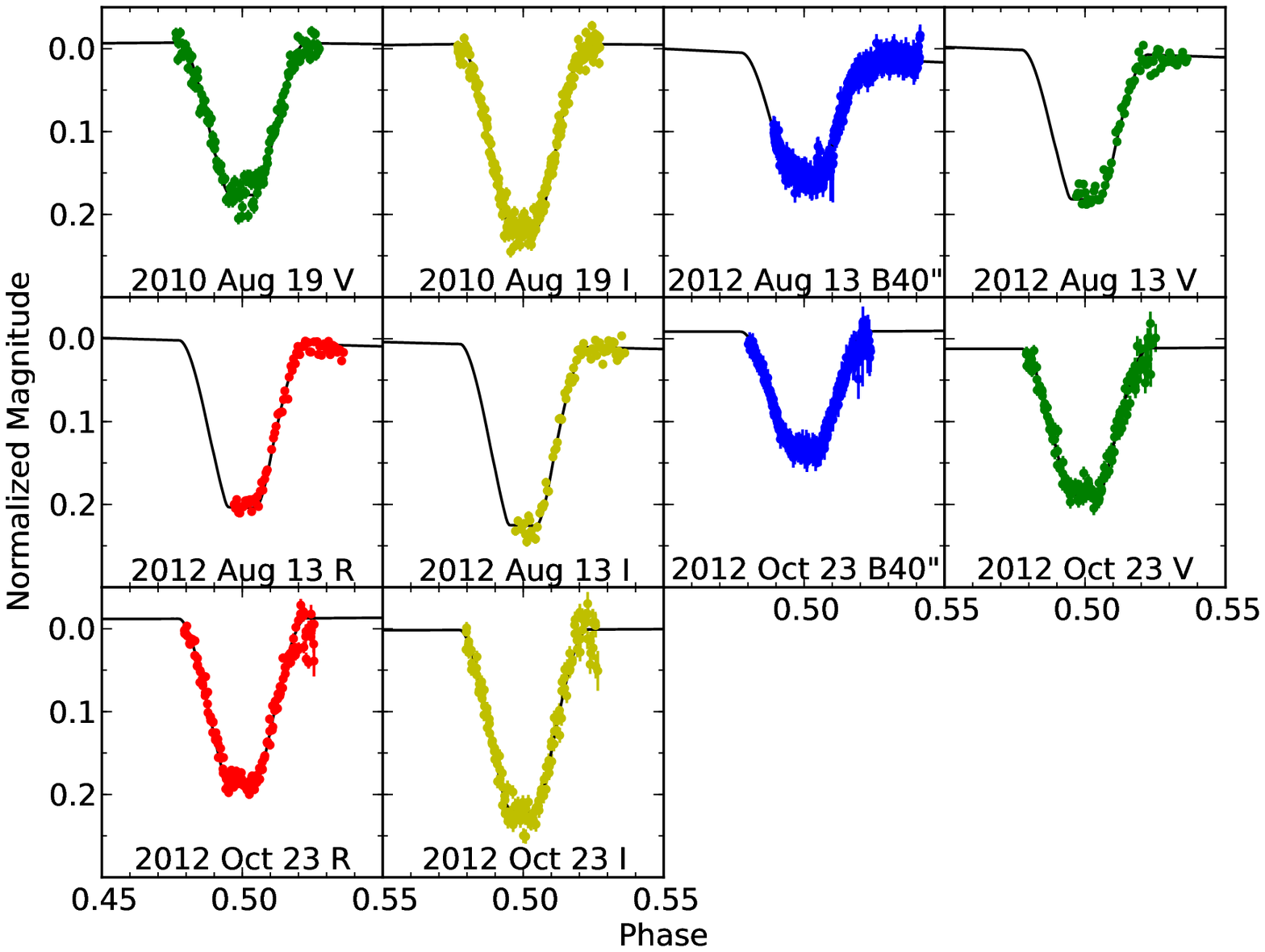}
\caption{Secondary eclipse light curves obtained from MLO observations in $B$ (blue points), $V$ (green points), $R$ (red points), and $I$ (yellow points) bands. The black curves show the ELC model that best-fits the photometry. As with the data in \autoref{fig:lcprim}, the light curves have been matched to the simultaneous \kep\ observations and thus the out-of-eclipse levels are not normalized to an instrumental magnitude of exactly zero.}
\label{fig:lcsec}
\end{figure*}
Due to the effects of stellar variability, the shape and depth of individual eclipses may vary from orbit to orbit. Consequently, the inferred stellar radii and temperatures measured from eclipses separated by several cycles may not agree. This statement remains true even if the model includes star spots, because we know that the star spot model is imperfect. In order to estimate the stellar parameters accurately (not just precisely), we split the light curve into sections and model them separately, and then examine the resulting set of system parameters. To be clear, ``separately'' is not ``independently'' since we use the global period, eccentricity, and contamination values obtained from analyzing the entire data set.

Each of the eight nights of ground-based photometry is paired with a section of the \kep\ light curve. The sections span 76--90\% of an orbital period, depending on the duration of the observations, and also because the star spots are expected to evolve on timescales longer than this. Unfortunately there is no \kep\ primary eclipse observation simultaneous with the 2010\,July\,22 observation, so the following secondary eclipse and out-of-eclipse star spot modulations are used. To emphasize the eclipses in the \kep\ light curve, rather than the out-of-eclipse spot variations (for which there are many more data points), we boosted the error bars by a factor of 3 for any datum that lies more than $\pm$0.05 in phase away from the eclipses. 

The ELC code was run on each section of data and the best-fit (lowest reduced chi-squared $\chi^{2}_{\nu}$) models are shown in \autoref{fig:nightkep} for the \kep\ data, in \autoref{fig:lcprim} for the MLO primary eclipses, and in \autoref{fig:lcsec} for the secondary eclipses. The parameter estimates are listed in \autoref{tab:res}, although the 16 spot parameter estimates for each night are omitted for brevity. As can be seen in \autoref{fig:lcsec}, the 1--2\,minute cadence ground-based photometry revealed that the secondary star is totally eclipsed for 0.01\,phase ($\sim$73\,minutes). On average, the star spots were $\sim$20\% cooler than the stellar temperature and were $\sim$15\,degrees in angular size. Although the fits to the eclipses are excellent, the relatively high values of the reduced $\chi^2$ are not surprising, given that the \kep\ uncertainties are so small and the star spot model is only a crude approximation. Our goal is not to measure the star spots or to obtain a perfect fit to the star spot-induced modulations in the light curve: we seek only to obtain accurate radius estimates. In the table we list only the best-fit value for the parameters, and not their uncertainty range, for the following reasons. For the night of 2010\,Jul\,22 we computed the formal 1$\sigma$ error bars in the standard way (determining the range spanned by the parameter of interest that corresponds with $\chi^{2}_{min} + 1$ relative to the best-fit model). This was computationally costly (so that repeating for all eight ground-based observations would require a great amount of time), but much more importantly, we found that the uncertainty estimates were significantly smaller than the spread of the best-fit parameters from night to night (underestimated by a factor of $\sim$2--10). In other words, the precision for a given night was much too high compared to the true uncertainty. Under the assumption that the uncertainties from each night are comparable to those of 2010\,Jul\,22, we determined the final best-fit parameters and the 1$\sigma$ errors from the unweighted average and standard deviation of the night-to-night best-fit solutions.

\begin{deluxetable*}{cccccccccc}
\tablecaption{Single Phase Orbital Properties of \kic\label{tab:res}}
\tablehead{\colhead{Date} & \colhead{$T_{\text{conj}}$} & \colhead{$M_1$} & \colhead{$q$} & \colhead{$R_1$} & \colhead{$R_1/R_2$} & \colhead{$T_{\text{eff,2}}$} & \colhead{$i$} & \colhead{$\chi_{\nu}^2$} \\ [-5pt] \colhead{} & \colhead{(BJD-2455000)} & \colhead{(\Msun)} & \colhead{} & \colhead{(\Rsun)} & \colhead{} & \colhead{(K)} & \colhead{(degrees)} & \colhead{}}
\startdata
2010\,Jul\,22   & -36.05682 & 0.985 & 0.792 & 1.312 & 1.65 & 5100 & 88.7 & 2.86 \\
2010\,Aug\,19 & -36.05703 & 0.974 & 0.794 & 1.304 & 1.62 & 5060 & 89.1 & 5.74 \\
2010\,Sep\,16 & -36.05665 & 0.993 & 0.796 & 1.314 & 1.63 & 5000 & 89.1 & 8.01 \\
2010\,Sep\,26 & -36.05683 & 0.983 & 0.794 & 1.309 & 1.63 & 5000 & 89.1 & 5.48 \\
2011\,Jul\,22   & -36.05682 & 0.985 & 0.793 & 1.313 & 1.65 & 5100 & 88.7 & 6.88 \\
2012\,Jul\,21   & -36.05695 & 0.986 & 0.793 & 1.303 & 1.61 & 5050 & 89.2 & 4.46 \\
2012\,Aug\,13 & -36.05683 & 0.981 & 0.797 & 1.306 & 1.62 & 5050 & 89.1 & 4.18 \\
2013\,Oct\,23  & -36.05600 & 1.005 & 0.781 & 1.324 & 1.61 & 5030 & 89.2 & 6.81
\enddata
\end{deluxetable*}

\subsection{Final Orbital Parameter Measurements}\label{sec:finparams}
\begin{figure*}
\plotone{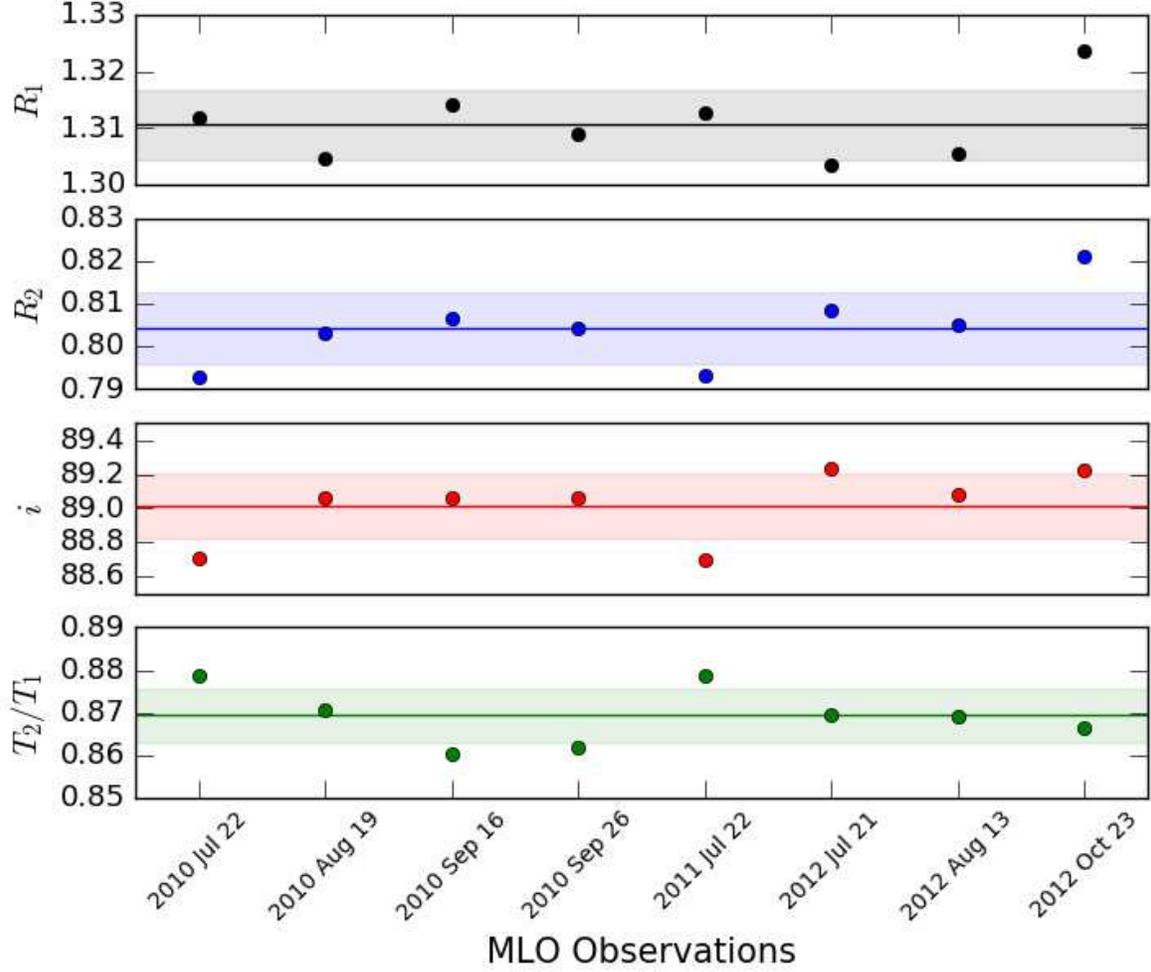}
\caption{The stellar radii, orbital inclinations, and temperature ratios determined at 8 different epochs. The dates indicate the nights of the ground-based observations. The means and standard deviations are show as the horizontal colored lines and shaded regions.}
\label{fig:rit}
\end{figure*}

\begin{figure*}
\epsscale{1.15}
\plottwo{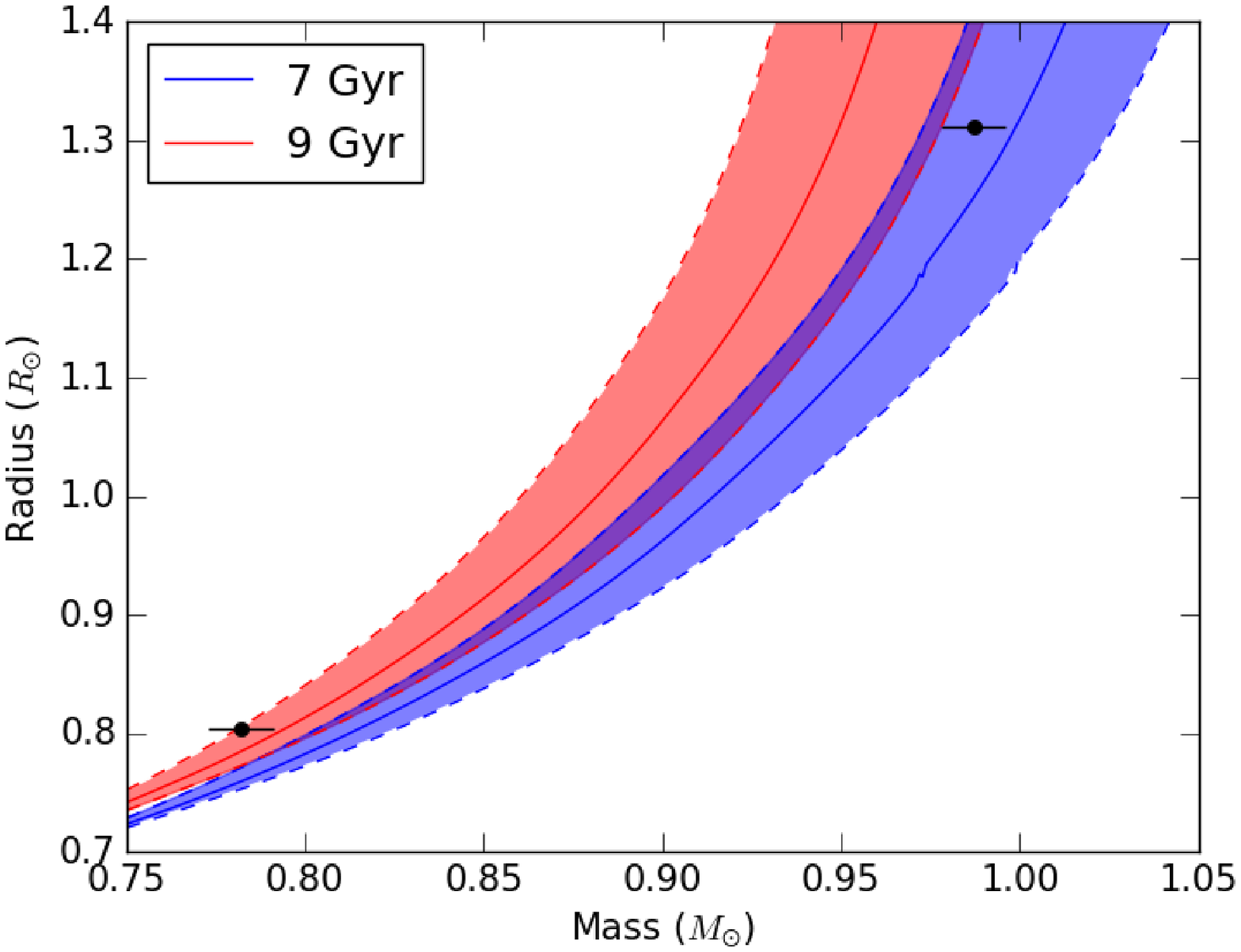}{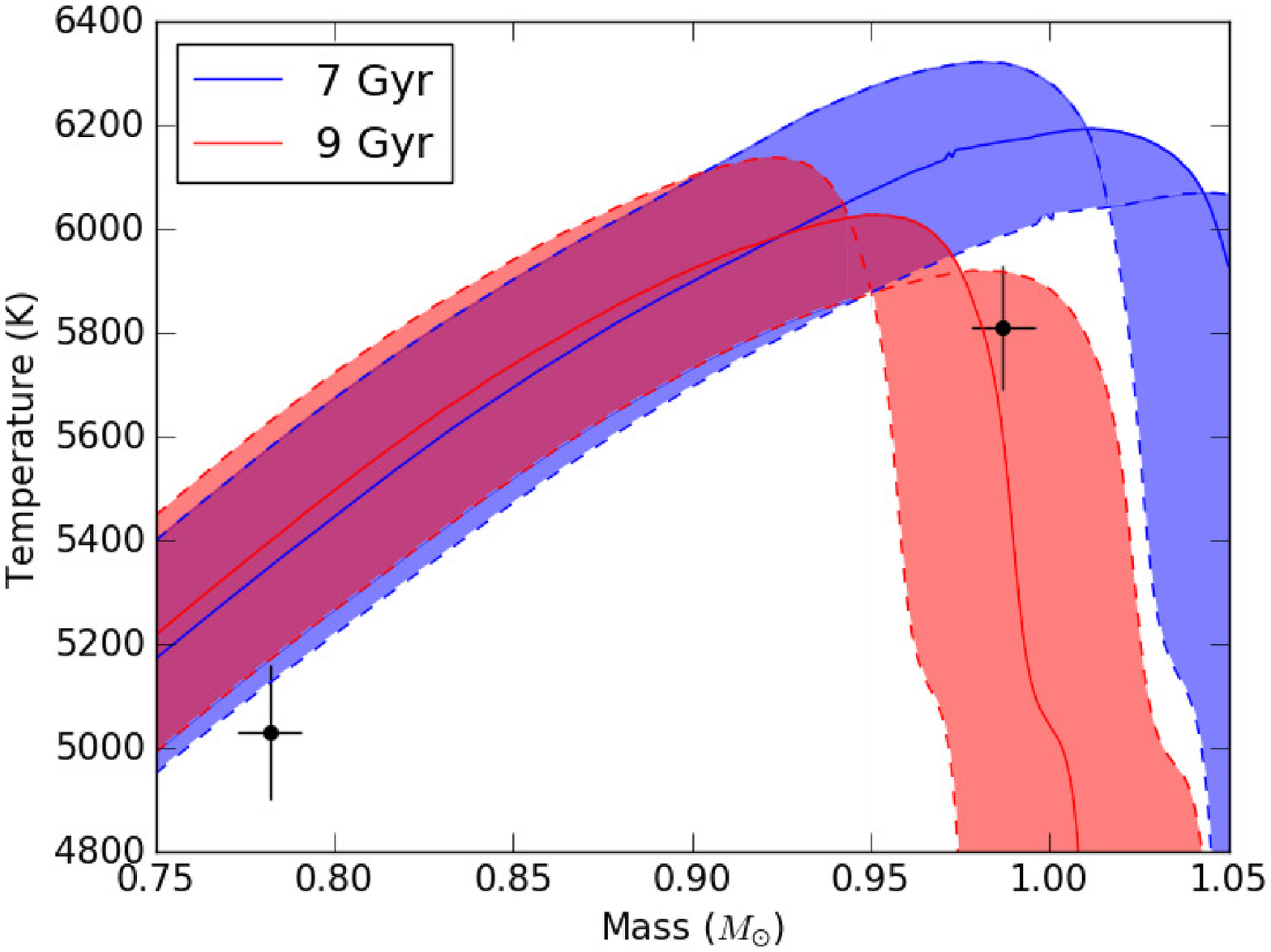}
\caption{The mass--radius (\textit{left}) and mass--temperature (\textit{right}) relationships for the stars in \kic. The solid curves in blue and red represent the 7\,Gyr and 9\,Gyr isochrones from the Dartmouth Stellar Evolution Program \citep{Dotter08} for a metallicity equal to [m/H]~$=-0.31$\,dex, respectively. The shaded regions between the dashed lines encompass the associated isochrones with assumed metallicities within the measured error of $\pm0.17$\,dex.}
\label{fig:mrt}
\end{figure*}

The final set of system parameters for \kic\ were determined from the unweighted average of the best-fit parameters derived from the eight simultaneous \kep\ and ground-based MLO light curves along with the radial velocity measurements. The 1$\sigma$ errors are defined by the standard deviation of the scatter in the eight estimates. These individual measurements, averages, and 1$\sigma$ uncertainties are shown for the stellar radii, system inclination, and stellar temperature ratio in \autoref{fig:rit}. The epoch-to-epoch changes emphasize the importance of multi-epoch observations and modeling for binary systems with active stars. The parameters and their respective errors are available in \autoref{tab:final}, although spot parameters are not included as they are variable with time. The measured masses ($0.987\pm0.009$\,\Msun\ and $0.782\pm0.009$\,\Msun) and radii ($1.311\pm0.006$\,\Rsun\ and $0.804\pm0.004$\,\Rsun) indicate the stars in \kic\ are G and K type stars, respectively. We use the best-fit stellar masses, radii, temperatures, and the spectroscopically measured metallicity ($\text{[m/H]}=-0.31\pm0.17$\,dex) to find the most suitable theoretical isochrones from the Dartmouth Stellar Evolution Program \citep{Dotter08}. We estimate the age of \kic\ to be 7--9\,Gyr and show the respective isochrones on the mass-radius and the mass-temperature diagrams in \autoref{fig:mrt}. It is unclear which stellar isochrone is most appropriate for \kic. There are several systems where at least one star does not match the stellar isochrones (e.g. UV Psc, \citealt{Popper97}; V636\,Cen, \citealt{Clausen09}; MG1-1819499, \citealt{Kraus11}; ASAS\,J065134-2211.5, \citealt{Helminiak19}; see \citealt{Feiden12}), but in the case of \kic, neither star exhibits a mass, radius, and temperature that is consistent with a single theoretical isochrone model. However, the Dartmouth stellar isochrone models \citep{Dotter08} assume solitary stars.

Regardless of the specific age, the primary star is consistent with being a Sun-like G star that is starting to leave the main sequence. The secondary star is a K star that is just within the population of stars that exhibit discrepant temperatures and radii \citep[$\lesssim$0.8\,\Msun, see][]{Torres10}. The temperature of the secondary star is $\sim$6\% cooler than predicted by either theoretical isochrone. The radius of the secondary star ranges from being consistent (assuming the 9\,Gyr isochrone) to up to 5\% larger (assuming the 7\,Gyr isochrone) than predicted by theoretical models, although stellar radius measurements may vary by up to 5\% due to stellar activity alone \citep{Feiden12}. 
\begin{deluxetable}{lc}
\tablecaption{System Parameters  for \kic\label{tab:final}}
\tablehead{\colhead{Parameter} & \colhead{Value}}
\startdata
$P_{\text{orb}}$ (days) & $5.069482 \pm 0.000001$ \\
$T_{\text{conj}}$ (BJD) & $2454963.9433 \pm 0.0003$  \\
$M_1$ ($\Msun$) & $0.987 \pm 0.009$ \\
$M_2$ ($\Msun$) & $0.782 \pm 0.009$ \\
$R_1$ ($\Rsun$) & $1.311 \pm 0.006$ \\
$R_2$ ($\Rsun$) & $0.804 \pm 0.004$ \\
$T_{\text{eff,2}}$ (K) & $5050 \pm 40$ \\
$T_{\text{eff,2}}/T_{\text{eff,1}}$ & $0.869 \pm 0.006$ \\
$i$ (degrees) & $89.0 \pm 0.2$ \\
$e$ & $<$10$^{-6}$ 
\enddata
\end{deluxetable}
%
%
%

% .................................................
\section{Periodic Signatures}\label{sec:periodicity}
Significant variations are observed in the out-of-eclipse light curve (see \autoref{sec:kepler}). When year-long segments of the \kep\ light curve were phase folded on the orbital period, we found that the resulting light curves were inconsistent with each other (see \autoref{fig:lcphase}), indicating that the stellar activity is significantly stronger than any permanent signals (e.g.,  Doppler beaming, ellipsoidal variations, or reflection effects). Under the assumption that the modulations are caused by star spots, the modulations can be used to measure the spin period(s) of the active star(s). In this section, we investigate periodic signatures of \kic\ to identify the stellar spins and, in effect, test theoretical stellar models given that old, circularized systems are expected to exhibit spin periods equal to the orbital period. We analyze the periodicities of the out-of-eclipse \kep\ light curve in \autoref{sec:kep_periods} and the residuals of the eclipse time measurements in \autoref{sec:omc}.

\subsection{Periodicities in the \kep\ Photometry}\label{sec:kep_periods}
\begin{figure}
\epsscale{1.15}
\plotone{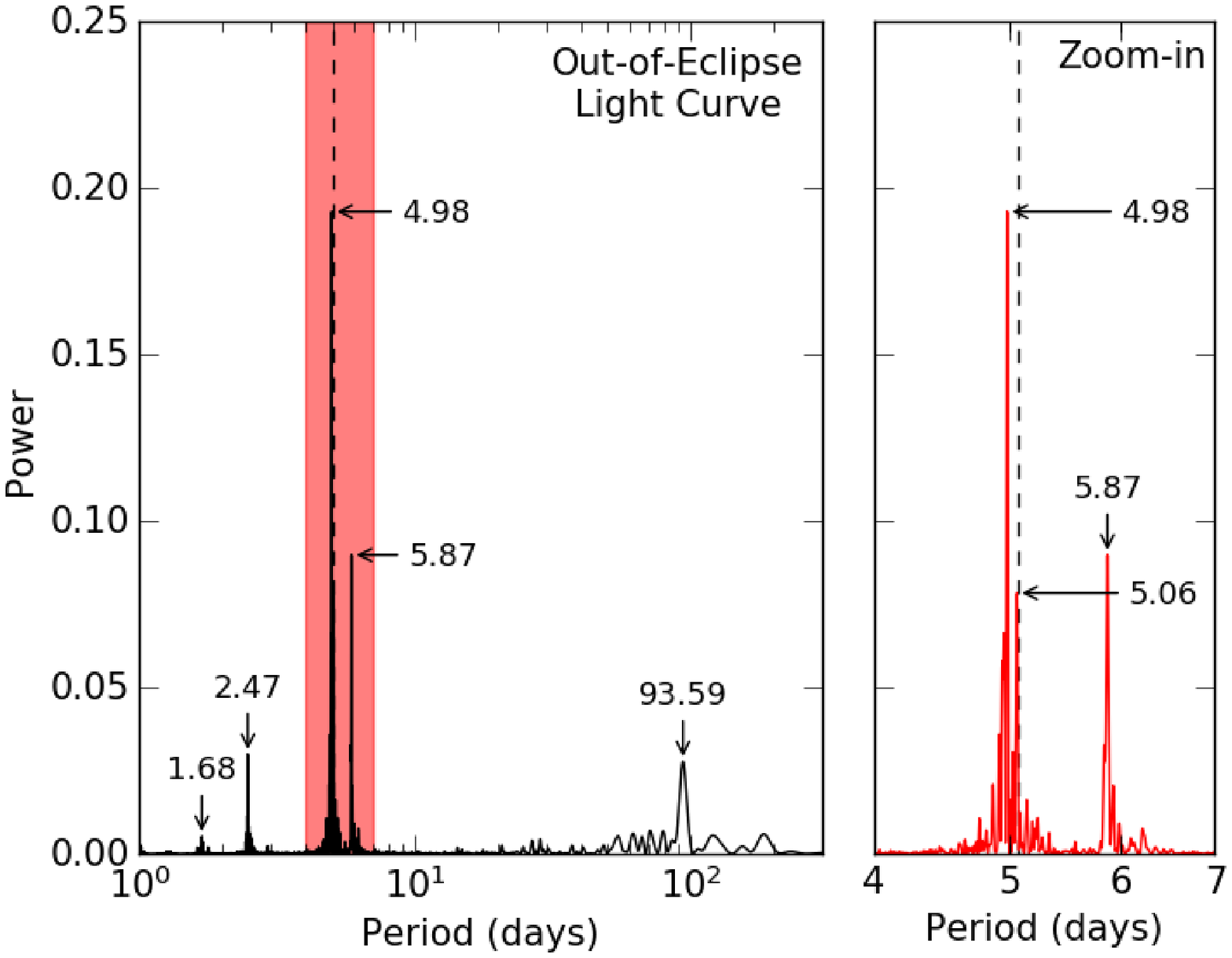}
\caption{Lomb-Scargle periodogram of the \kep\ light curve after eclipses have been cut for \kic. The periods of significant peaks are noted and marked with arrows. \textit{Left:} The full periodogram from 1--300\,days. \textit{Right:} A zoom-in of the shaded region between 4--7\,days, emphasizing the structure in the periodogram. The vertical dashed lines mark the orbital period of \kic.}
\label{fig:periodogram}
\end{figure}
In \autoref{fig:periodogram} we show the Lomb-Scargle periodogram \citep{Lomb76, Scargle82} of the entire \kep\ light curve with eclipses removed. We find several strong peaks near the  orbital period (5.07\,days) and a peak at 93.59\,days. The latter is likely an artifact of the \kep\ Quarterly roll and CCD change, so we ignore it. The power near the orbital period is much more interesting, and consists of three distinct peaks at $\sim$4.98\,days, $\sim$5.87\,days, and $\sim$5.06\,days. The 5.06\,day spike is consistent with the orbital period, and as such suggests that even though the eclipses have been removed, there is some phenomenon that exists at the timescale of the orbital period, such as low-amplitude Doppler beaming hidden by the star spot modulations. The largest peak, at $\sim$4.98\,days, is likely due to star spots on the primary star and is consistent with the spectroscopically measured rotational velocity within the uncertainties. Therefore, 4.98\,days is the rotation period of the star. The star spot modulation hypothesis is supported by the presence of weaker, but significant, spikes at the 2nd and 3rd harmonics. The $\sim$5.87\,day periodicity is tentatively associated with the spin of the secondary star, though harmonics are not seen in this case and is 1.1$\sigma$ slower than the spectroscopically measured rotational velocity. If these identifications are correct, it is a very peculiar situation: the primary star is spinning super-synchronously while the secondary is spinning sub-synchronously. Both of these are surprising given that the system has a circular orbit ($e<10^{-6}$) and has had plenty of time ($\sim$8\,Gyr) to synchronize the spins with the orbital period (timescale for synchronization is $\ll$1\,Gyr). \citet{Zahn89} stated that circularization occurs pre-main sequence for binaries with orbital periods $<$8\,days, but they will be spun-up when they reach the zero age main sequence due to contraction. In extension, \citet{Zahn94} adds that stars spin down with age as they lose angular momentum through stellar winds, causing the spin periods to be less than the orbital periods. Using models of angular momentum evolution for binary systems, \citet{Keppens00} found that decay of the orbital separation occurs rapidly when one of the component stars leaves the main sequence, but that the spins should remain synchronized or nearly synchronized as the orbital period changes. We speculate that the peculiar rotations may be related to angular momentum exchange within the binary system as the primary star evolves off the main sequence. 

Interestingly, the $\sim$5.87\,day spike is ``cleaner'' than the $\sim$4.98\,day spike, meaning there is less power extending out as tails from the base of the peak. Since both periodicities suffer from the same sampling window, this difference is probably not due to a difference in leakage into sidelobes. The excess power near the base of the $\sim$4.98\,day spike is possibly due to the evolution of the star spots (changing amplitudes and random phasing), or perhaps more interestingly, due to star spots at different latitudes coupled with differential rotation \citep[e.g.,][]{Lurie17}. 
\begin{deluxetable}{lcc}
\tablecaption{Significant Lomb-Scargle Periodogram Peaks \label{tab:periodogram}}
\tablehead{\colhead{Identification} & \colhead{Period} & \colhead{Equation}}
\startdata
Orbital Period & 5.069482 & $P_{\text{orb}}$ \\ 
Candidate Spin 1 & 4.980 & $P_{\text{spin,1}}$ \\ 
2nd Harmonic & 2.470 & (1/2)$\times P_{\text{spin,1}}$ \\ 
3rd Harmonic & 1.680 & (1/3)$\times P_{\text{spin,1}}$ \\ 
Beat Period 1 & 280.140 & $[(P_{\text{spin,1}})^{-1}-(P_{\text{orb}})^{-1}]^{-1}$ \\ 
Candidate Spin 2 & 5.869 & $P_{\text{spin,2}}$ \\ 
Beat Period 2 & 37.253 & $[(P_{\text{orb}})^{-1}-(P_{\text{spin,2}})^{-1}]^{-1}$ \\ 
Quarter Gap & 93.588 & ~ \\ 
~ & 98.436 & ~ \\ 
Unknown & 19.534 & ~
\enddata
\end{deluxetable}

\subsection{Periodicities in the O--C Diagram}\label{sec:omc}
\begin{figure}
\epsscale{1.15}
\plotone{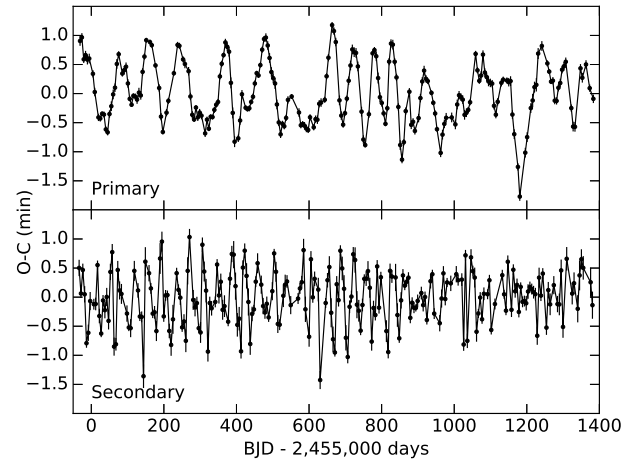}
\caption{The O--C eclipse timing variations versus time for the primary (top panel) and secondary (bottom panel) eclipses, respectively.}
\label{fig:omc}
\end{figure}
\begin{figure*}
\epsscale{1.15}
\plottwo{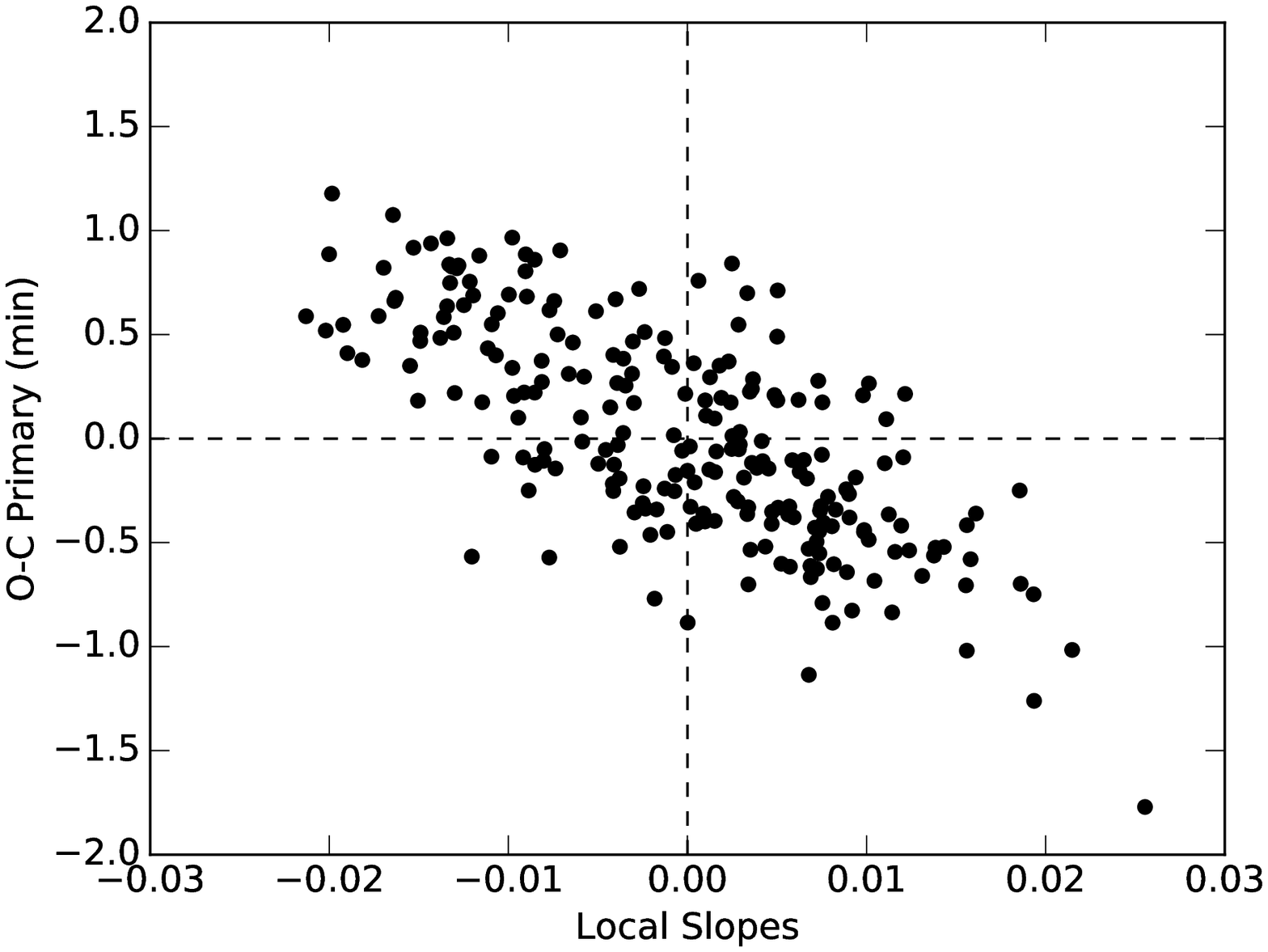}{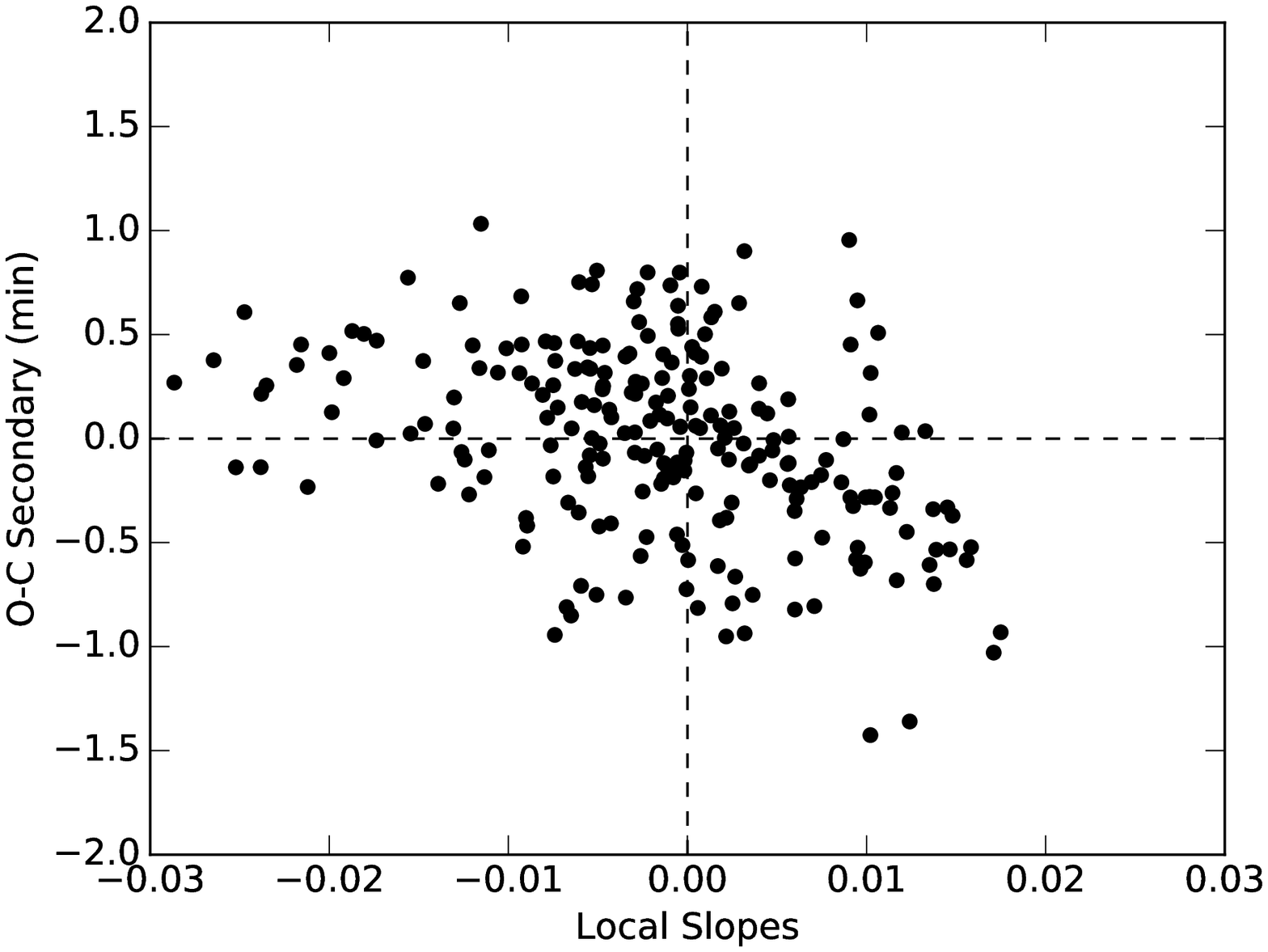}
\plottwo{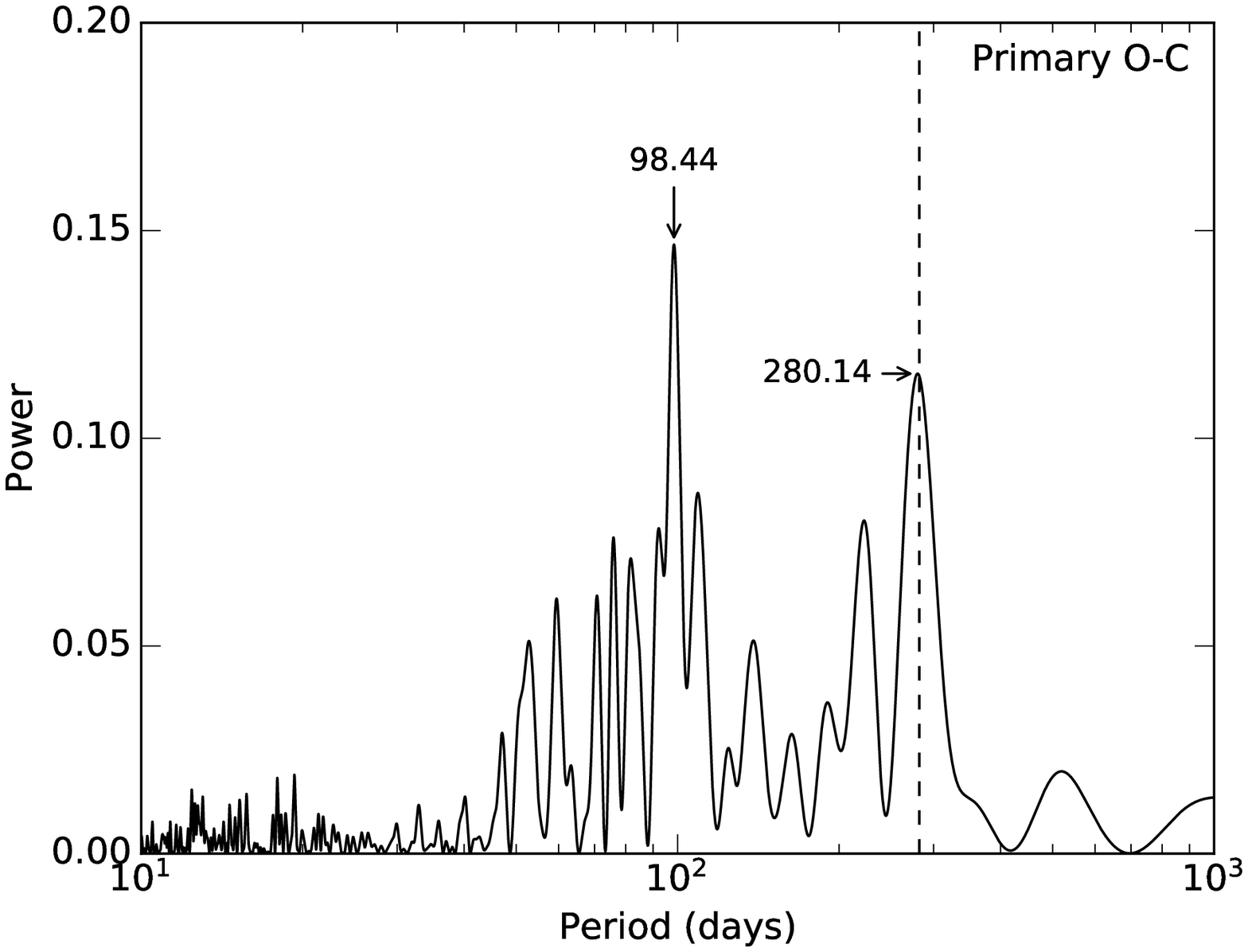}{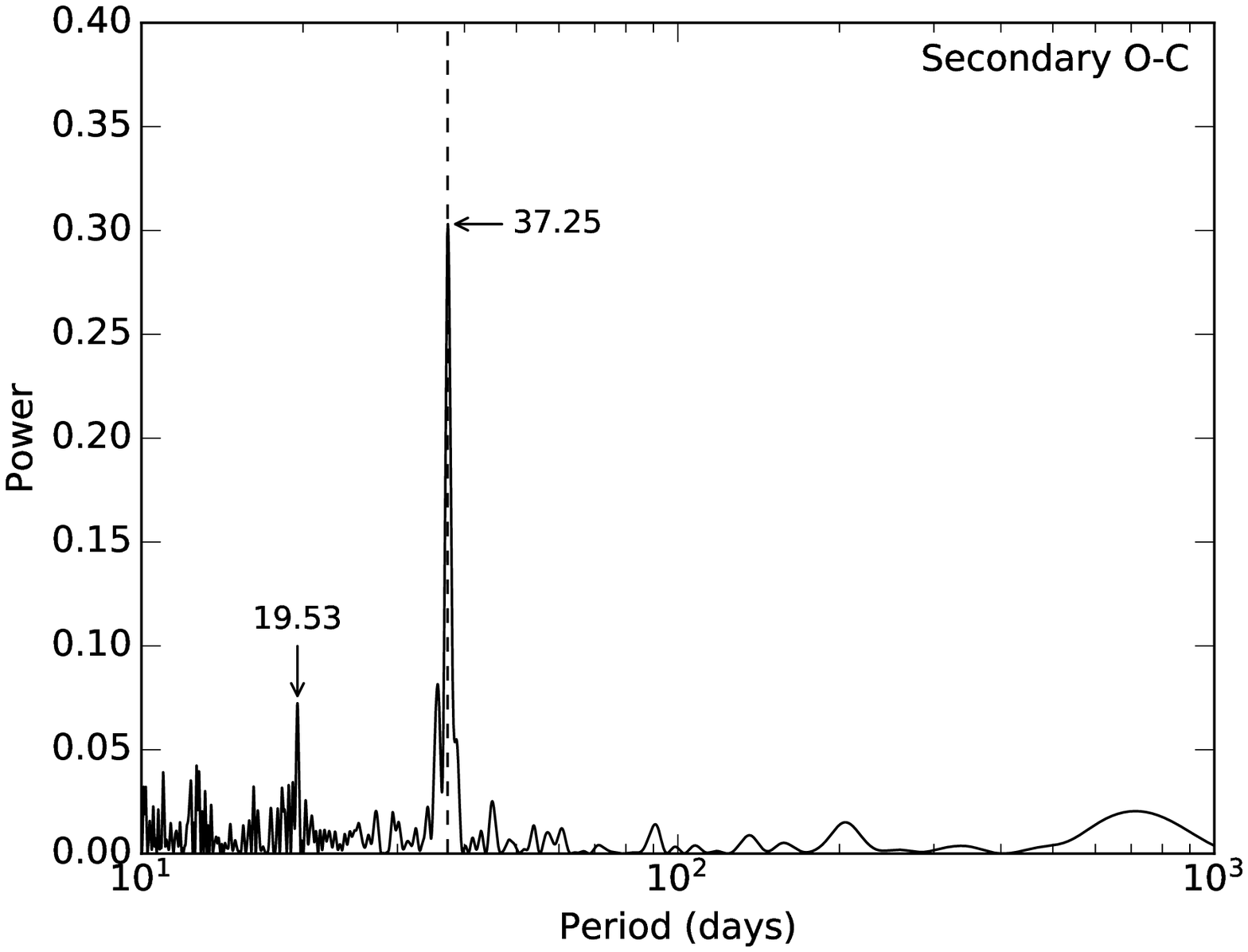}
\caption{\textit{Top:} The O--C primary (\textit{left}) and secondary (\textit{right}) eclipse timings versus the local slopes of the \kep\ light curve. \textit{Bottom:} Lomb-Scargle periodograms of the O--C primary (\textit{left}) and secondary (\textit{right}) eclipse timings. The vertical dashed lines mark the expected beat periods between the orbital period and the stellar spins.}
\label{fig:omc_periodogram}
\end{figure*}

If the orbital motion of the binary stars were purely Keplerian, the mid-eclipse times would be separated by intervals of the orbital period. The ``observed minus calculated'' (O--C) diagram would then be flat within the observational noise. However, gravitational perturbations (light-travel time effect, classical and relativistic apsidal motion, dynamical effects) can cause the O--C diagram to exhibit curvature and other variations. In addition, star spots can shift the apparent times of the eclipses, causing spurious variations in the O--C diagram. For example, if a star spot is on the side of the stellar disk rotating into view, the light curve will be decreasing (negative slope as the star spot becomes more apparent), and the center of the eclipse profile will be shifted towards later times (because covering a star spot will cause the ingress of the eclipse profile to be higher in flux). This is assuming that the spin is prograde with respect to the orbital motion \citep{Mazeh15}. In \autoref{fig:omc} we show the O---C diagrams for the primary and secondary star eclipses, and while generally flat, they clearly show significant fluctuations. We can be certain that some of these variations are due to star spots by plotting the deviations from zero in the O--C against the local slope of the light curve surrounding each eclipse. If the variations were caused by gravitational interactions, there should be no correlation. But if the variations are caused by star spots, then we expect the eclipse times to be shifted to later times as the star spot comes into view and to earlier times as the star spot rotates out of view behind the star's limb. This behavior is clearly seen in the top left panel of \autoref{fig:omc_periodogram}, as there is a strong (negative) correlation between the shift in time and slope of the light curve.

To better understand the eclipse timing variations, we compute the Lomb-Scargle periodogram of the O--C time series. These are also shown in the bottom panels of \autoref{fig:omc_periodogram}. The periodogram of the primary star eclipse times is essentially just noise, with the most significant peak being consistent with the \kep\ Quarterly roll. The periodogram of the secondary star eclipse times shows a very prominent spike at $\sim$37.25\,days. The origin of this spike is hypothesized to be a beat between the orbital period (5.069482\,days) and the spin period of the secondary star ($\sim$5.869\,days). These yield a period of 37.21\,days, consistent with the observed signal. An equivalent explanation is that the $\sim$37.25\,day signal is an alias of the $\sim$5.869\,day spin period that has been mirrored from above to below the Nyquist frequency of the binary. Similarly, these explanations can be extended to the second most significant peak in the periodogram of the primary star eclipse times at $\sim$280.14\,days, as it is consistent with the expected beat period between the orbital period and the spin period of the primary star ($\sim$4.980\,days) of 282.14\,days. Finally, there remains a weak $\sim$19.534\,day periodicity that is unexplained; it is not a harmonic of the $\sim$37.25\,day period. All of the significant peaks and their identified source are listed in \autoref{tab:periodogram}.

\section{Summary}\label{sec:summary}
We combined 4\,years of continuous space-based observations from the \kep\ space telescope with radial velocity measurements and 8 epochs of ground-based photometry from Mount Laguna Observatory to solve for the orbital and stellar parameters of \kic. We find that \kic\ is an edge-on ($i=89^\circ$) double-lined eclipsing binary with a 5.07\,day circular orbit. The primary star has a mass and radius equal to 0.99\,\Msun\ and 1.31\,\Rsun, respectively, making it an old, Sun-like star that is beginning to leave the main sequence. The secondary star is a K\,star with a mass and radius equal to 0.78\,\Msun\ and 0.80\,\Rsun, respectively, and it exhibits a temperature that is $\sim$6\% cooler than predicted by theoretical models. 

Our analysis demonstrates that the when the stars are significantly spotted, it is important to analyze independent eclipses in order to properly estimate the uncertainties in the system parameters. With only a single high quality light curve, one might unintentionally confuse accuracy with precision and obtain underestimates of the uncertainties. This work emphasizes the importance of measuring orbital and stellar properties of active stars from multiple epochs, when possible.

There are 1--4\% peak-to-peak fluctuations in the out-of-eclipse \kep\ light curve caused by stellar activity. From analyzing the \kep\ light curve and O--C eclipse timing variation periodograms, we propose candidate spin periods of approximately 4.98\,days and 5.87\,days for the primary and secondary stars. Neither of these periods match the orbital period, and thus neither star is synchronously rotating. In fact, the primary appears to be spinning super-synchronously while the secondary is rotating sub-synchronously---a puzzle. There are a number of reasons why a star may be sub-synchronously rotating, such as angular momentum loss from stellar winds, orbital separation decay as the primary star evolves, angular momentum conservation as the primary star evolves, or differential rotation. But a super-synchronous spin is more challenging to explain. We leave a more in-depth investigation of the stellar rotation characteristics to future studies, as they may benefit from future observations from ongoing space missions such as the \textit{Transiting Exoplanet Survey Satellite}. \kic\ is also known as TIC\,164527774 and will be observed in \textit{TESS} sectors 14 and 26. Overall, \kic\ proves to be an interesting case study for testing stellar astrophysical models due to the evolutionary state of its primary star and asynchronous spins. 
%
%
%

% ..................................................................
\acknowledgments
We thank Trevor A. Gregg, Carolyn Heffner, Aja Canyon, Michael Hill, and Beverly Thackeray-Lacko for obtaining MLO observations. TF acknowledges support from the Minority Biomedical Research Support---Initiative for Maximizing Student Development (MBRS-IMSD) program, which was funded by the National Institutes of Health/National Institute of General Medical Sciences (NIH/NIGMS) grant \#5R25GM058906-12. The authors also gratefully acknowledge support from the National Science Foundation via grants AST-1109928 and AST-1617004. WFW and JAO thank John Hood, Jr.\ for his generous support of exoplanet research at SDSU. The HET is a joint project of the University of Texas at Austin, the Pennsylvania State University, Stanford University, Ludwig-Maximilians-Universit\"{a}t M\"{u}nchen, and Georg-August-Universit\"{a}t G\"{o}ttingen. The Hobby-Eberly Telescope is named in honor of its principal benefactors, William P. Hobby and Robert E. Eberly. This paper includes data collected by the \kep\ mission. Funding for the \kep\ mission is provided by the NASA Science Mission directorate.

\facilities{HET (HRS), Kepler, MLO:0.6m (SBIG STL-1001E CCD), MLO:1m (2005 Lorel CCD)}
\software{AstroML \citep{VanderPlas12}, Astropy \citep{Astropy_collaboration13, Astropy_collaboration18}, ELC \citep{Orosz00}, IRAF \citep{Tody86, Tody93}, Matplotlib \citep{Hunter07}, NumPy \citep{Oliphant07}, SciPy \citep{Oliphant07}}

% ..................................................................
% \bibliography{KIC8736bib}
% \bibliographystyle{aasjournal}

\end{document}